\documentclass{aastex62}

\received{January 1, 2018}
\revised{January 7, 2018}
\accepted{January 20, 2019}
\submitjournal{ApJ}

\shorttitle{The MRI imprint on the Short-GRB Jets}
\shortauthors{Sapountzis \& Janiuk}

\begin{document}
\title{The MRI imprint on the Short-GRB Jets
}

\correspondingauthor{A. Janiuk}
\email{agnes@cft.edu.pl}



\newcommand{\lrz}{\gamma}
\newcommand{\enth}{\xi}
\newcommand{\hslope}{h}
\newcommand{\polind}{\hat \Gamma}



\author{Konstantinos Sapountzis$^{1}$}
\author{Agnieszka Janiuk$^{1}$}
\affil{$^{1}$
Center for Theoretical Physics, Polish Academy of Sciences, 
Al. Lotnikow 32/46, 
02-668 Warsaw, Poland}
\email{kostas@cft.edu.pl}

\begin{abstract}
Short gamma ray bursts are presumably results of binary neutron star mergers, which lead to the formation of a stellar mass black hole, surrounded by a remnant matter. The strong magnetic fields help collimate jets of plasma, launched along the axis of the black hole rotation. We study the structure and evolution of the accreting plasma in the short GRBs and we model the formation of the base of a relativistic, Poynting-dominated jets. Our numerical models are based on the general relativistic MHD, axisymmetric simulations. We discuss the origin of variability in the GRB jet emission, which timescales are related to the action of the magneto-rotational instability in the accreting plasma. We also estimate the value of a maximum achievable Lorentz factor in the jets produced by our simulations, and reached at the large distances, where the gamma ray emission is produced.
\end{abstract}



\section{Introduction}

\indent Almost two decades have passed after the realization that the compact object binaries, namely the Black Hole-Neutron Star (BHNS) and the Neutron Star-Neutron Star (NSNS) binaries \citep{Eichler_Livio_Piran_schramm_1989_Nature, Paczynski_1991_Acta_Astronomica, Narayanetal1992}, are eligible progenitors for the Short Gamma Ray Bursts (sGRB). The associated complexities on both macroscopical and microphysical properties of the outflow result to a still ongoing effort to identify crucial aspects of the central engine operation. Some of the fundamental requirements for the mechanism responsible for the outflow launching were already stated years ago. The cosmological origin of the bursts infers a spectacular release of energy, $10^{52}-10^{54}$ ergs for isotropic emission or two orders of magnitude less \citep{Fongetal2015}, when a collimated outflow is considered and the prompt radiation flux is corrected for the beaming factor $f_b=\left(1-\cos\theta_j\right)$, where $\theta_j$ is the jet half opening \citep{Sarietal1999, Rhoads1999}.

\indent The minimum variability time scale (MTS) can be used together with the causality principle to provide a rough estimation of the emitting region. The measurement of MTS is a challenging task, since the GRBs present intense variability in the subsecond regime and the signal distribution to the various time-bins is distorted by the white-noise of the observation. Nevertheless \citet{MacLachlan2013} used a wavelet analysis adjusted for the GRB phenomenology and provided the variability timescale of the shorter bursts, $\left(10^{-4}-10^{-2}\right)s$ and an order of magnitude higher for the long-duration ones. Up today, there is no definite conclusion for the origin of the observed variability. The leading models, the internal-shock scenario \citep{Narayanetal1992, ReesMeszaros1994}, and/or the photospheric emission (see \citet{Granot2015} for review), link the variability directly to the properties of the central engine. In the long duration bursts the situation can be further complicated by the external environment interference because of the flow propagation in the interior of the collapsar progenitor \citep{woosley1993, woosleyheger2006, pernaetal2018}. Notice though that in the short bursts the environment of propagation is expected to be of lower density. In the long gamma ray bursts, the variability may also originate from the internal, current-driven instabilities, see \citep{2003astro.ph.12347L}.

\indent Independently of the process responsible for the variability, the inferred high energy density of the emitted prompt radiation photons, together with the observed non-thermal $\gamma$-ray spectra consists the so called \textit{compactness problem}. According to it, the enormous optical depth due to the electron-positron pair production produces a thermalized emission rather than the observed non-thermal one. The compactness problem was resolved on the theoretical grounds by invoking relativistic expansion with a rather large bulk Lorentz factor, $\lrz>10^{2}$ \citep{Paczynski_1986, Baring_1997}. In its turn, the acceleration of the outflow in this high Lorentz factor requires exceedingly clean explosions with ejecta masses of $<10^{-5}M_{\odot}$, a condition known in the literature as the baryon-loading problem \citep{Shemi_1990, Lei_2013}.

\indent The precise nature of the compact binary progenitor has not been fully determined yet. A major breakthrough occurred recently with the multi-messenger detection of the GRB 170817A through the VIRGO/LIGO and Fermi-GBM observatories. The waveform of the emitted gravitational waves is consistent with the merging of a system of two neutron stars. Moreover, it was possible to set limit on the system members, and total masses under a $90\%$ credible interval,  $m_1 = \left(1.81\pm0.45\right) M_\odot$, $m_2=\left(1.11\pm0.25\right)M_\odot$ and $M_{\rm tot}=2.82^{+0.47}_{-0.09}M_\odot$ for a high spin prior restriction or $m_1 = \left(1.48\pm0.12\right) M_\odot$, $m_2=\left(1.26\pm0.10\right)M_\odot$ and $M_{tot}=2.74^{+0.04}_{-0.01}M_\odot$ for a low spin prior restriction model, respectively \citep{Abbott2017}. The event was followed by a sGRB after $\sim 1.7 s$, defined as the time between the merging instant and the Fermi Gamma-ray Space Telescope trigger time. Nevertheless, the electromagnetic component of the phenomenon was reported as few orders fainter than a typical short burst, $E_{\rm iso}=\left(3.1\pm0.7\right)\cdot 10^{46}$ ergs and of course the poor statistics, since we have only one event, does not exclude the possibility that the BHNS mergers serve also as a distinguished class of progenitors in the general sGRB context. The off jet-axis observation and the line of sight through the surrounding cocoon \citep{Lazzati2017}, a structured jet \citep{2018MNRAS.473L.121K}, or an intrinsic property of the specific NSNS merging event \citep{Murguia_Berthier2017, Zhang_et_al_2017arXiV} can both provide a plausible interpretation for the fainter radiation. The statistics themselves are expected to be improved as the time pases and the gravitational wave detectors improve their sensitivity providing more robust conclusions for the nature of the merging system.

\indent From the theoretical point of view, the current study of the merging phase, the last spiral orbits of the binary members and the ability of the central engine to launch an outflow, is performed through full general relativistic simulations, i.e. ones that evolve also the space-time \citep{BaiottiRezolla2016, Paschalidis_2017}. Assuming more or less realistic configurations for the initial magnetic field and the equation of state of the neutron star/stars involved a scheme to handle the neutrino transport. These type of simulations consist the current state of the art. Beyond the difficulties of the underlying physical process, they also have to face the large scattering on the spatial scales involved in the phenomenon. 
Among the potential remnants, a massive NS and the long lasting supra-massive neutron star (SMNS) do not lead to the formation of a BH-torus system that powers the bursts \citep{Shibata2000, Margalit2015}. On the contrary, a hyper-massive neutron star (HMNS) can conclude in a BH-torus engine, if the living time of the differential rotating NS is short enough, while the same holds for the prompt collapse to a BH. In fact, the  delayed collapse of a HMNS might be the reason for the observed difference between the merging and GRB launching instants on the GW 170817 event \citep{Granot2017}, while the early optical and later infra-red emission due to r-process argue also in favor of the HMNS scenario \citep{MargalitMetgzer2017}.

\indent Although the proper framework for the merging of a compact object binary is the full GR simulation, the integrations performed on a fixed space-time, namely the Kerr one, provide a useful insight for the underlying processes resulting to the jet launching. A number of phenomena like the magnetic barrier formation \citep{Bisnovatyi_1976, McKinney_2012}, the Blandford-Znajek efficiency \citep{Blandoford_Znajek_1977} and the implication of the neutrino emission on the outflow \citep{Justetal2016} can be easier identified using a fixed space-time, 2D or 3D, algorithmic scheme. Nevertheless, the interpretation of both the accretion process and the jet acceleration phase by a single scheme is still missing. The extended range of the spatial scales involved and the hyper-relativistic Lorentz factors achieved are two obstacles requiring adaptive mesh refinement and other novel techniques, e.g. the use of GPU \citep{Liska2018}, no matter of which type of the space-time evolution is adopted. The extended spatial scales become even more crucial for Poynting dominated outflows, since the magnetically driven acceleration generally acts at larger distances than the thermally driven one \citep{Granot2015}. 

\indent The aim of the present work is to investigate the origin of the intense variability of the prompt $\gamma-$ray observations in conjunction with the conditions applying at the accreeting torus. We bypass the spatial scale problem by focusing on the magnetic energy flux along the field line, instead of the bulk Lorentz factor. This quantity is central on the special relativistic MHD theory of jets, and provides also the maximum achievable Lorentz factor. Its accurate evolution can be specified though only through an analytic or numerical scheme and when the profile of the external pressure has been specified.

\indent In our model we assume an ideal conducting flow and a realistic dipolic magnetospheric configuration, i.e. one that is extended beyond the interior of the NS remnant. The main focus of our study is the imprint of the torus magnerotational instability, MRI \citep{Balbus_Hawley_1991}, on the time variability of the energy flux of the emerging jet which is performed through a set of numerical simulation using the GRMHD code \textit{HARM} \citep{Gammie_2003, Noble_et_all_2006}. In general, any configuration where the magnetic energy dominates much over the rest mass energy is a challenging task for most available numerical schema. It poses significant restrictions to initial magnetic field magnitude because of the induced $\beta-$parameter of the interstellar plasma, defined as the ratio of the thermal to the magnetic pressure, $\beta \equiv p_{\rm g} / p_{\rm mag}$. The specific type of limitations are further enhanced by the formation of the magnetic barrier resulting to rather low $\beta$ flows in the jet-funnel, even if the material in the interior of the torus is thermally dominated. The specific choice of the magnetic field configuration is not new. For example, one of the first full GRMHD simulations adopted a similar configuration and concluded to an electromagnetically driven jet outflow \citep{Paschalidis_Ruiz_Shapiro_2015}. Their integration resulted in a two order order of magnitude enhance of the BH magnetic field above the poles through the field line winding and for the ISM $\beta-$parameter range $\beta_{\rm ISM}=10^{-2}-10^{-1}$, resulting to the launch of a mildly relativistic jet, $\gamma_j\sim 1.3$, for $100$ ms, with Poynting luminosities $10^{51}$ erg s$^{-1}$, and magnetization values $\sigma\equiv B^2/\left(8\pi\rho c^2\right) \sim 100$ which are consistent with the typical short GRBs. A crucial component of the simulation, as of any electromagnetic launching model, is the presence of a poloidal field component that is dictated by the jet launching conditions \citep{Beckwith_2008}. 

\indent Beyond the Blandford-Znajek process which extracts the rotational energy of the BH, the neutrino annihilation is an alternative/complementary process that taps the gravitational energy of the disk \citep{Mochkovitch_et_al_1993, Aloy_et_al_2005, Janiuk2013, Liu2015, Janiuk2017}. Although our algorithmic schema is able to handle a more realistic EOS including neutrino emission, for the current model we chose to focus on the purely electromagnetic effects and ignore the potential neutrino implications on the launching process. Since the magneto-rotational instability is the leading phenomenon in our model, the limiting lower value of the magnetic field was restricted by the condition that the instability is properly resolved. A characteristics of the specific instability is the independence of the maximum growth rate from the magnitude of the magnetic field, and its dependence from the rotational velocity of the disk. As a consequence, we examine two sets of models with different initial $\beta-$parameter for the torus configurations, while every set includes models differing with the radius where the maximum of the pressure and the magnetic field maximum dominance occurs.
The results prove the correlation between the outflow time variability and the maximum growth rate of the instability. 

\indent The article is organized as follows. In Section 2 we present the initial configuration of our model and we investigate the restrictions under which the MRI is properly resolved. The results of the integration appear in Section 3, while the analysis and conclusions are the subject of Section 4.

\section{The Simulations Setup}

\indent We use the general relativistic magnetohydrodynamic code, \textit{HARM} \citep{Gammie_2003, Noble_et_all_2006}, to integrate our model under a fixed Kerr metric, i.e. neglecting effects like the self gravity of the  disrupted material, the BH spin acceleration and the system orbiting around the center of mass. The \textit{HARM} code is a finite volume, shock capturing scheme that solves the hyperbolic system of partial differential equations, once brought in conservative form, by implementing a Harten, Lax, van Leer schema (HLL) to calculate numerically the corresponding flux function. In terms of the Boyer-Lindquist coordinates, $\left(r,\theta,\phi\right)$, the black hole is located at $0<r \leq r_{\rm hor}$, where $r_{\rm hor} = \left(1+\sqrt{1-a^{2}}\right) r_g$ is the horizon radius of a rotating black hole with mass $M$ and angular momentum $J$ in geometrized units, $r_g=G M /c^2$, and $a$ is the dimensionless Kerr parameter, $a=J/(Mc), 0 \leq a \leq 1$. In our simulations we investigate the jet launching from highly rotating black holes, $a=0.9$ and corresponding  horizon at $r_{\rm hor}=1.46 r_g$. Our simulations are axisymmetric, and the computational domain extends in $r-\theta$ direction.

\subsection{The Torus Initial Configuration}

\indent The remnant of the accreting material is modeled following \citet{Fishbone_Moncrief_1976_ApJ} who provided an analytic solution of a constant angular momentum, steady state (thereafter FM). Other similar configurations with a power law radial evolution of the angular momentum \citep{Chakrabarti1985}, or of independently varying Bernoulli parameter (sum of the kinetic energy, potential energy, and enthalpy of the gas) and disk thickness \citep{Pennaetal2013} are also possible. In the FM model, the position of the material reservoir is determined by the radial distance of the innermost cusp of the torus, $r_{\rm in}$, and the distance where the maximum pressure occurs, $r_{\rm max}$.
Because of its geometry, the relative difference of the two radii determines also the dimension of the torus, with higher differences resulting to extended cross section. Subsequently the $r_{\rm in}$ and $r_{\rm max}$ determine also the angular momentum value and the distribution of the angular velocity along the torus, a crucial parameter for the magnetrotational instability. Following \citet{Gammie_2003}, we use the value of the maximum initial density of the torus to scale the density value over the integration space
\begin{eqnarray*}
r_g=\frac{G M}{c^2}=1.48\cdot 10^5 \frac{M}{M_\odot} \rm{cm} \\
t_g=\frac{r_g}{c}=4.9\cdot 10^{-6} \frac{M}{M_\odot} \rm{s} \\
f_g= 6.0 \cdot 10^{20}\sqrt{\rho_{\rm torus}} \left(\frac{M}{M_\odot}\right)^2 \rm{Mx}
\end{eqnarray*}
where $r_g, t_g, f_g$ are the spatial, time and magnetic flux units, respectively. The last scaling is derived by the cgs definition of the magnetic units and the $r_g, t_g, \rho=\rho_{\rm torus} \, \rm{[g\,cm^{-3}]}$ units are assumed. By construction, the FM model describes the steady state hydrodynamical fluid around a Kerr hole and as a consequence the magnetized configurations considered here are not in equilibrium. Nevertheless as long as the plasma $\beta-$parameter remains high we might assume that the fluid is close to steadiness.

\indent The $r_{\rm in}$ and $r_{\rm max}$ radii of the torus are related to a number of factors that is very difficult to determine in advance. For example, in the BHNS progenitor the radii are related to the distance where the tidal disruption of the neutron star occurs which in turn depends on its equation of state, while they are also affected by the NS magnetic field strength and topology and by the self-rotation of the binary system's members. An even more complicated situation takes place in the NSNS type of progenitors. There is a significant number of fully relativistic simulations estimating this distance, most of which give a value from a few to few tens of geometrical radii \citep{Ferrari_2010, Rezzolla_et_al_2011, Ruiz2016, Paschalidis_Ruiz_Shapiro_2015, Kumar_2017}. In our work we use the fiducial radius of $r_{\rm max} = (12-25)\,r_g$ to identify the implications of MRI on the emerged jet time variability, while we also included a rather large value, $r_{\rm max} = 50\,r_g$, to examine the parametrical extension of our model. We also assumed a polytropic equation of state, $p_g = K \rho^{\polind}$, with $\polind=4/3$ , $K=10^{-3}$.

\indent The magnetic field configuration resembles the magnetic field produced by a circular current. The form of such field can be easily found in a number of text books, e.g. \citep{jackson_classical_electrodynamic}
\begin{eqnarray}
  A_{\phi}({r,\theta})= A_0 \frac{\left(2-k^2\right) K\left(k^2\right)-2 E\left(k^2\right)}{k\sqrt{4 R r \sin\theta}} \\
  k = \sqrt{\frac{4 R r \sin\theta} { r^2 + R^2 + 2 r R \sin\theta}} \nonumber
\end{eqnarray}
where $E,K$ are the complete elliptic functions and $A_0$ is a constant that it is used to scale the magnetic field and the initial $\beta$-parameter across the initial torus.

\indent Our magnetic field configuration is different than usually adopted by the authors of the fixed spacetime general relativistic 3-D simulations of jets, e.g., \citet{2018MNRAS.tmp.2798F}. Nevertheless in the full GR simulations it is being used as a \citet{Paschalidis2013} prescription for the NS involved in the merger. In our case, the extended magnetic field outside the NS describes the existence of magnetic field in the intermediate space between the two merging objects, while it can also describe the potential BH magnetic field created by the HMNS collapse (magnetic braking process).

\subsection{The Magneto-Rotational Instability}

\indent One of the main difficulties in simulating the ideal conducting magnetized accretion is the resolution of the magnetorotational instability. 
By construction, the accreting reservoir is enriched by the magnetic field that resembles the field of a circular conductor with radius equal to the radius of the maximum pressure of the torus. Since the magneto-rotational instability, MRI \citep{Balbus_Hawley_1991}, has a key role in our model, the limiting lower value of the magnetic field was restricted by the condition that the instability is properly resolved. As a consequence, a second set of models describing highly thermally dominated flows attained values of  $\beta \sim \rm{few}$ hundreds. A characteristic of the specific instability is the independence of the maximum growth rate from the magnitude of the magnetic field, but its dependence from the rotational velocity of the disk. Thus a further parametrization of our configurations was performed in terms of the maximum pressure radius of the FM torus. 

\indent The maximum growth rate $\Omega_{max}$ for a disk in Kerr metric is obtained by \citet{Gammie_2004} and for an observer at infinity it is given by
\begin{equation}
  \Omega_{\rm max}^2=\frac{9}{16} \Omega^2 \frac{D^2}{C}
  \label{eq:growthrate}
\end{equation}
where $D=1-2/r+(a/r)^2, C=1-3/r+2 a r^{-3/2}$ are the \citet{Novikov_Thorne_1973} correction parameters and $\Omega$ is the rotational velocity of the fluid. 

\indent Similarly to the Newtonian case, the growth rate does not depend on the magnetic field magnitude. In contrast, the wavelength of this mode does, and for sufficiently low values of the magnetic field it becomes unmanageably small by a numerical scheme, setting a lower limit on the potential $A_0$ values (an upper limit for $\beta$). Following \citet{Noble_2010, Hawley_2011, Narayan_et_al_2012}, we introduced the MRI resolution parameter along the $\theta$-direction
\begin{equation}
 Q_{\rm MRI}^\theta = \frac {2 \pi v_{A\theta}}{\Omega d\theta}
\end{equation}
where $v_{A\theta}=b^\theta / \sqrt{\left(\rho \enth \right)^2 + b^\mu b_\nu}$ is the $\theta$-component of the Alfv\'{e}n velocity and $d\theta$ is the grid cell size as measured in the fluid frame. At the limit of the vertical field, $Q_{\rm MRI}^\theta$ tends to $\lambda_{\rm MRI} / d\theta$, where $\lambda_{\rm MRI}$ is the wavelength of the fastest growing mode. 

\indent The task of determining the resolution able to describe properly the nonlinear coupling between the various modes is a rather complicated one and its study was performed mostly via numerical simulations. In their work, \citet{Hawley_2011} investigated a number of both shearing box and global simulations and examined a set of diagnostics. Following their suggestion, all the models of the high $\beta$-parameter we present attain meridional values $Q_{\rm MRI}^\theta \ge 10$, while the models with lower $\beta$-parameter are well beyond that limit with $Q_{\rm MRI}^\theta \ge 100$.

\subsection{Initial Parameters}

\indent The extended configuration of the magnetic field enriches the significantly lighter interstellar medium (ISM) with a magnetic content and induces a much lower $\beta$-parameter. As a consequence, the ISM follows its own evolution at the initial steps of integration and depending on the magnetic content it might accrete, form an outflow, or even kick the torus away if the outflow is strong enough. In order to keep the ISM $\beta$-parameter in the range of our code capabilities and in a density magnitude modeling a rarefied exterior environment, a specific lower bound of the torus initial $\beta$ was chosen. As a consequence, at the initial stages of our simulation part of the ISM is accreted seeding the black hole with a parabolic shape magnetic field without disrupting the system.

\indent The \textit{HARM} code doesn't perform the integration in the Boyer-Lindquist coordinates, but instead in the so-called Modified Kerr-Schild ones: $\left(t,x^{(1)},x^{(2)},\phi\right)$ \citep{Noble_et_all_2006}. The transformation between the coordinate systems is given by:

\begin{eqnarray*}
r= R_0 + \exp\left[{x^{(1)}}\right]
\\
\theta = \frac{\pi}{2} \left(1 + x^{(2)}\right) + \frac{1 - \hslope}{2} \sin\left[\pi\left(1 +x^{(2)}\right)\right]
\end{eqnarray*}
where $R_0$ is the innermost radial distance of the grid, $0 \leq x^{(2)} \leq 1$, and $\hslope$ is a parameter that determines the concentration of points at the mid-plane. In our models we use $h=0.3$ (notice that for $h=1$ and a uniform grid on $x^{(2)}$ we obtain an equally spaced grid on $\theta$, while for $\hslope=0$ the points concentrate on the mid plane). The exponential grid in the $r$-direction leads to a higher resolution and it is adjusted to resolve the initial propagation of the jet outflow. Summarizing, the selected grid resolution is $1020\times512$. Moreover, due to the significant acceleration of our outflow is imposed, it was necessary to consider a 'gamma-barrier' and test different values to exclude any implications on the launched jet variability. The value $\gamma_{\rm max} = 50$, common for all models, was chosen such that it doesn't affect the grid regime that our study is concentrated on.

\indent The density floor, i.e. the magnitude of minimum density allowed by our numerical scheme, is assumed on the order of $10^{-8}$ with respect to the normalized density in code units and correspondingly the internal energy floor at $10^{-10}$ code units  \citep{Penna}. The floor is triggered in conjunction with a minimum $\beta$ floor, $\beta_{\rm min} \sim 10^{-4}$, which is determined by the ability of our code to approach the force free limit. Beyond that limit numerical errors are induced and the code fails in the specific cells adding an interpolation process from the neighboring the cells primitive variables values. The choice of the $\beta_{\rm min}$ floor was made as to avoid this process in the jet-funnel and for the domain of interest.

\begin{figure}
    \centering
     \includegraphics[width=0.45\textwidth]{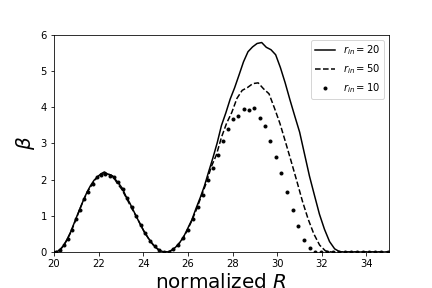}
     \includegraphics[width=0.45\textwidth]{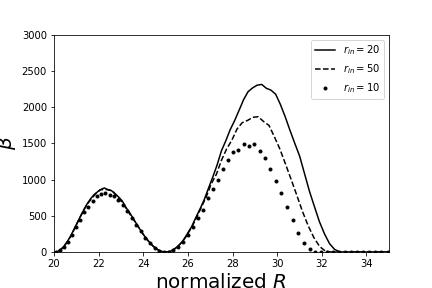}
     \caption{The initial distribution of the plasma-$\beta$ at the equator and inside the torus. \textbf{Left:} models with higher magnetic content. \textbf{Right:} models with a lower one. In both diagrams the solid line stands for the models indicated by \textit{HD}, the dashed for the \textit{MD} and the dashed-dot for the \textit{LD}. The normalization on the radial scale is made as all of the models has similar radial size and center at the same point as the \textit{MD} models. Notice that the parameter of the \textit{LD-Therm} model differs by a factor of $\sim 2$, so that the MRI resolution criterion is satisfied.} 
     \label{fig:in_beta}
\end{figure}

\begin{deluxetable}{lcccccc}

\tablecaption{Summary of the models \label{tab:in}}

\tablehead{
  \colhead{Model} 
  & \multicolumn{2}{p{3cm}}{\centering  Torus radii \\ $r_g$ units \\ $r_{in} \qquad\;\;\;\; r_{max}$}
  & \multicolumn{1}{p{3cm}}{\centering ISM density $\rho$ (code units} 
  & \multicolumn{1}{p{2cm}}{\centering $A_0$ \\ $f_g$ units} 
  & \multicolumn{1}{p{2cm}}{\centering $T_{MRI}$ \\ $t_g$ units} 
  &  \multicolumn{1}{p{2cm}}{\centering $Q_{MRI,min}^\theta$}}

\startdata
      HD-Therm & 50 & 60 & $1.6 \cdot 10^{-9}$ & $10.0$ & $630$ & $10$ \\
      HD-Magn & 50 & 60 & $8.6 \cdot 10^{-8}$ & $200.0$ & $630$ & $151$ \\
      MD-Therm & 20 & 25 & $1.0 \cdot 10^{-8}$ & $1.6$ & $174$ & $10$\\
      MD-Magn & 20 & 25 & $3.9 \cdot 10^{-7}$ & $32.0$ & $174$ & $173$\\
      LD-Therm & 10 & 12 & $4.0 \cdot 10^{-8}$ & $0.16$ & $61$ & $13$\\
      LD-Magn & 10 & 12 & $2.5 \cdot 10^{-7}$ & $3.1$ & $61$ &  $122$  \\
\enddata
\tablecomments{The first two models refer to a distant torus, while the rest ones refer to models more suitable for a realistic sGRB configuration. The dimensional quantities above are expressed in the geometrized units as being further specified by the density normalization (see main text). The characteristic time of the fastest growing mode and the minimum value of the $Q_{MRI}^\theta$ parameter across the meridional plane appear respectively in the last two columns.}
\end{deluxetable}

\indent Table~\ref{tab:in} presents summary of the models and their initial parameters, and the expected characteristic time of the MRI calculated from Eq.~\ref{eq:growthrate}, as well as the minimum instability resolution across the meridional plane.
The models are labeled with HD, MD, and LD symbols, which stand for the high, medium, and low size of the torus, respectively. The further distinction between the 'Therm' and 'Magn' types of simulations is based on the dominance of the thermal, and magnetic pressure, as prescribed by the magnitude of plasma $\beta$ in the initial conditions (see Fig. ~\ref{fig:in_beta}). 
The plasma-$\beta$ parameter is not constant throughout the torus. Its equatorial distribution is shown in Figure~\ref{fig:in_beta}. Note that the \textit{LD-Therm} model deviates from the rest of thermal models in order to satisfy the $Q_{\rm MRI,min}^\theta$ condition.

\indent In our approach, we use the plasma energy-momentum tensor $T^{\mu\nu}$
\begin{eqnarray*}
{T_{\left(m\right)}}^{\mu\nu}= \rho \enth u^\mu u^\nu + p g^{\mu\nu} \\
{T_{\left(em\right)}}^{\mu\nu}=b^\kappa b_\kappa u^\mu u^\nu+\frac{1}{2} b^\kappa b_\kappa g^{\mu\nu} - b^\mu b^\nu\\
T^{\mu\nu}={T_{\left(m\right)}}^{\mu\nu}+{T_{\left(em\right)}}^{\mu\nu}
\end{eqnarray*}
where $\enth$ is the fluid specific enthalpy, to introduce the magnetization parameter $\sigma$ and the energy parameter $\mu$
\begin{equation}
 \sigma= \frac{\left(T_{em}\right)^r_t}{\left(T_{m}\right)^r_t} 
 \qquad
 \mu = - \frac{T^r_t}{\rho u^r}
\label{eq:mu}
\end{equation}
Assuming the flat space-time the physical interpretation of the above quantities corresponds respectively to the Poynting energy flux towards $\hat r$ normalized to the thermal, plus inertial energy flux of the matter and to the total plasma energy flux normalized to the mass flux (see \citet{Vlahakis_2003a, Komissarov2009}). Nevertheless, close to the BH the physical interpretation stops to be valid and the $\mu,\sigma$ are interpreted as two mathematical quantities. 

\indent The motivation of our notation is based on the special relativistic theory of the magnetized steady state jets. In the special relativistic framework, the energy conservation along a field line is expressed by $\mu=\lrz \enth \left(1+\sigma\right)$ as the sum of the inertial-thermal energy of the plasma, $\lrz \enth$, and its Poynting flux, $\lrz \enth \sigma$. One can readily verify that $\mu$ quantity determines also the maximum achievable Lorentz factor $\lrz_{\infty} = \mu$, when all the Poynting and the thermal energy is transformed to baryon bulk kinetic form ($\sigma \to 0 , \enth \to 1$). But the conditions under which a highly accelerated outflow is obtained is a complicated issue depending on both the initial conditions applying at the base of the outflow and the overall topology of the magnetic field configuration. Using the conclusion of \citet{Vlahakis_2003a}, and especially their cold rotator model, the closest to the force free outflow, the acceleration due to the magnetic content takes place when the line's cylindrical arm, $\varpi$, becomes $\log_{10}(\varpi/\varpi_0) \sim 2$ orders higher from its initial value, $\varpi_0$. For example, on a field line originating from $\varpi_0\sim 10$ that point corresponds to $\varpi\sim 1000$ and as a consequence to a vertical distance where our $r$-logarithmic grid becomes too spare.

\section{Results}

\subsection{General structure of the outflow}

\begin{figure}
    \centering
     \includegraphics[width=0.2\textwidth]{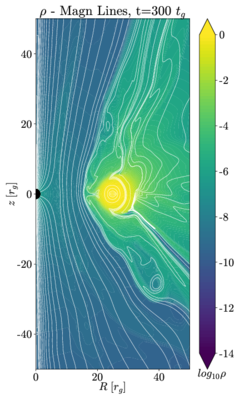} 
     \includegraphics[width=0.2\textwidth]{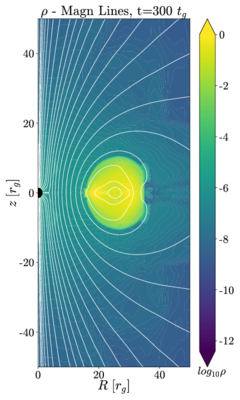} 
     \caption{The torus evolution in an early integration time $t = 300 t_g$ for the $r_{in} = 20 r_g$ models. \textbf{Right:} In the \textit{Therm} models the torus keeps its form, although some accretion is already initiated. \textbf{Left:} In the \textit{Magn} models the higher magnetic pressure results to the dilation and deformation of the torus.} 
     \label{fig:in_phase}
\end{figure}

\indent Our simulations are divided in two sets with the \textit{Magn} class of models referring to a weakly magnetized torus and the \textit{Therm} class to torus with a significantly higher $\beta$-parameter. The overall evolution follows a similar pattern, no matter the type of model we investigate. Since the interstellar medium is not in equilibrium there is an initial phase where the ISM acrretes, seeding BH with a parabolic magnetic field which results in the early formation of the magnetic barrier. The torus itself might be close to a steady state retaining its structure for a long integration time (\textit{Therm} models), or further away from equilibrium leading to a dilated formation (\textit{Magn} models, see Figure~\ref{fig:in_phase}).

\begin{figure*}
    \centering
     \includegraphics[width=0.8\textwidth]{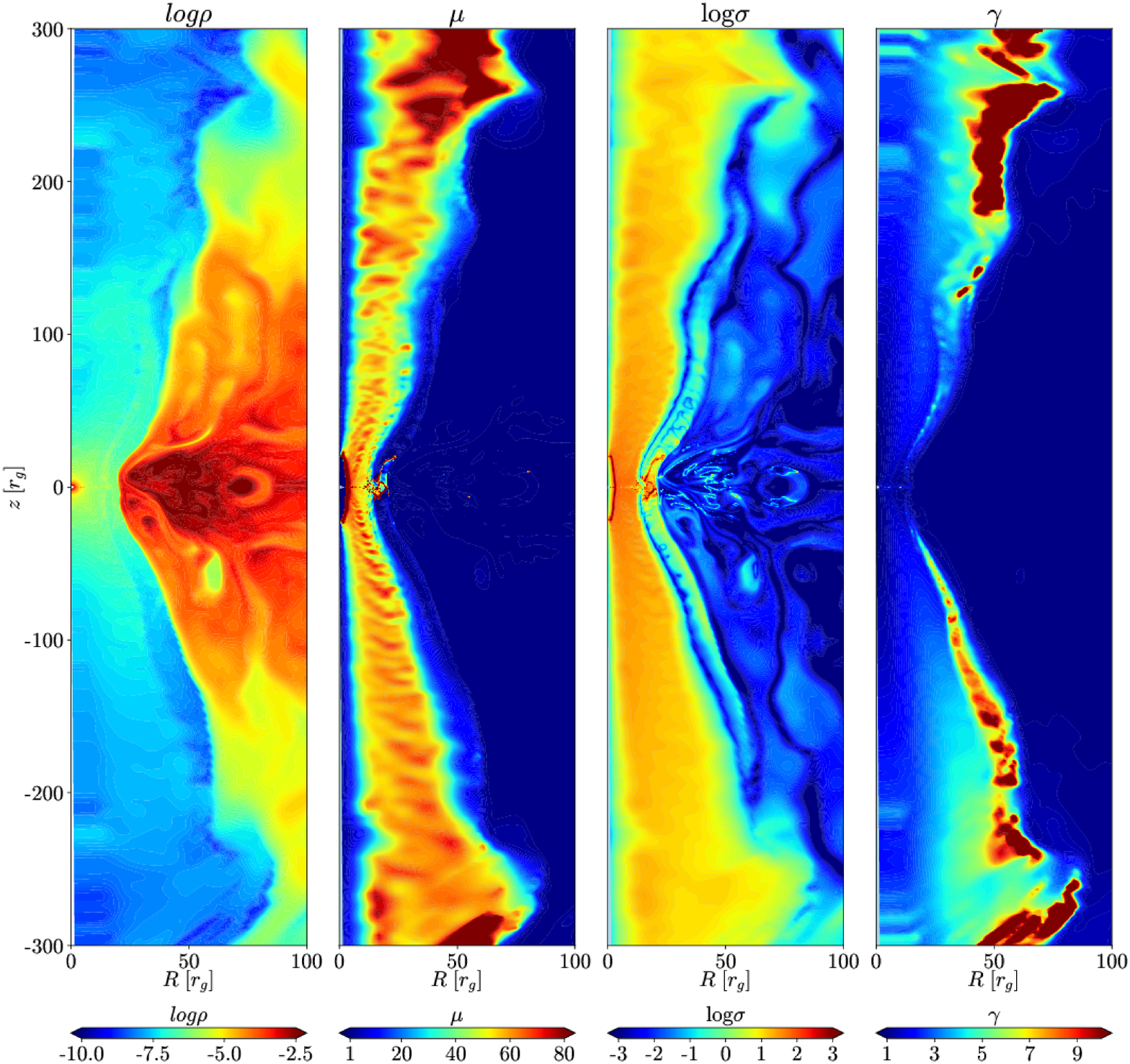}
     \caption{The $t=2000t_g$ snapshot of the \textit{MD-Magn} model. \textbf{1st panel:} The spatial distribution of the density logarithm, the low baryon jet has lower density by 3-4 orders of magnitude from its surrounding environment.
The value of density just after the horizon is $9.5\times10^{-4}$.
       \textbf{2nd panel:} The energetic parameter, $\mu$, indicates that most of the energy outflow occurs inside the jet funnel. \textbf{3rd panel:} The logarithm of the magnetization parameter exhibiting a Poynting-dominated jet, $\sigma>10$. \textbf{4th panel:} The acceleration and the Lorentz factor achieved is only mildly relativistic with significant acceleration $\lrz>10$ only at the exterior boundary of the jet.}   
     \label{fig:combined_snapshot}
\end{figure*}

\indent The formation of a magnetic barrier leads to the launch of a low baryon jet in all the models considered. Using  \textit{MD-Magn} model as a fiducial model, Figure~\ref{fig:combined_snapshot} shows the $t=2000t_g$ snapshot and some characteristic quantities of the outflow ($\rho , \mu , \sigma, \lrz$). Two regions, the jet funnel and its surrounding matter-dominated environment, are clearly distinguished. The dominant energetic flux is contained in the jet funnel as a Poynting-dominated outflow with magnetization parameter of a few to few tenths. The jet funnel is also the area where most energy is released with values $\mu>50$, suitable to describe the sGRBs outflows. The acceleration of the outflow is not significant and beyond the boundary the rest of the flow remains only mildly relativistic. The acceleration in shorter spatial scales of the outer part of the outflow is associated with significant thermal content induced on the outflow at the base and the effects of the magnetic barrier at the boundary of the outflow. 

 \begin{figure*}
    \centering
    \includegraphics[width=0.23\textwidth]{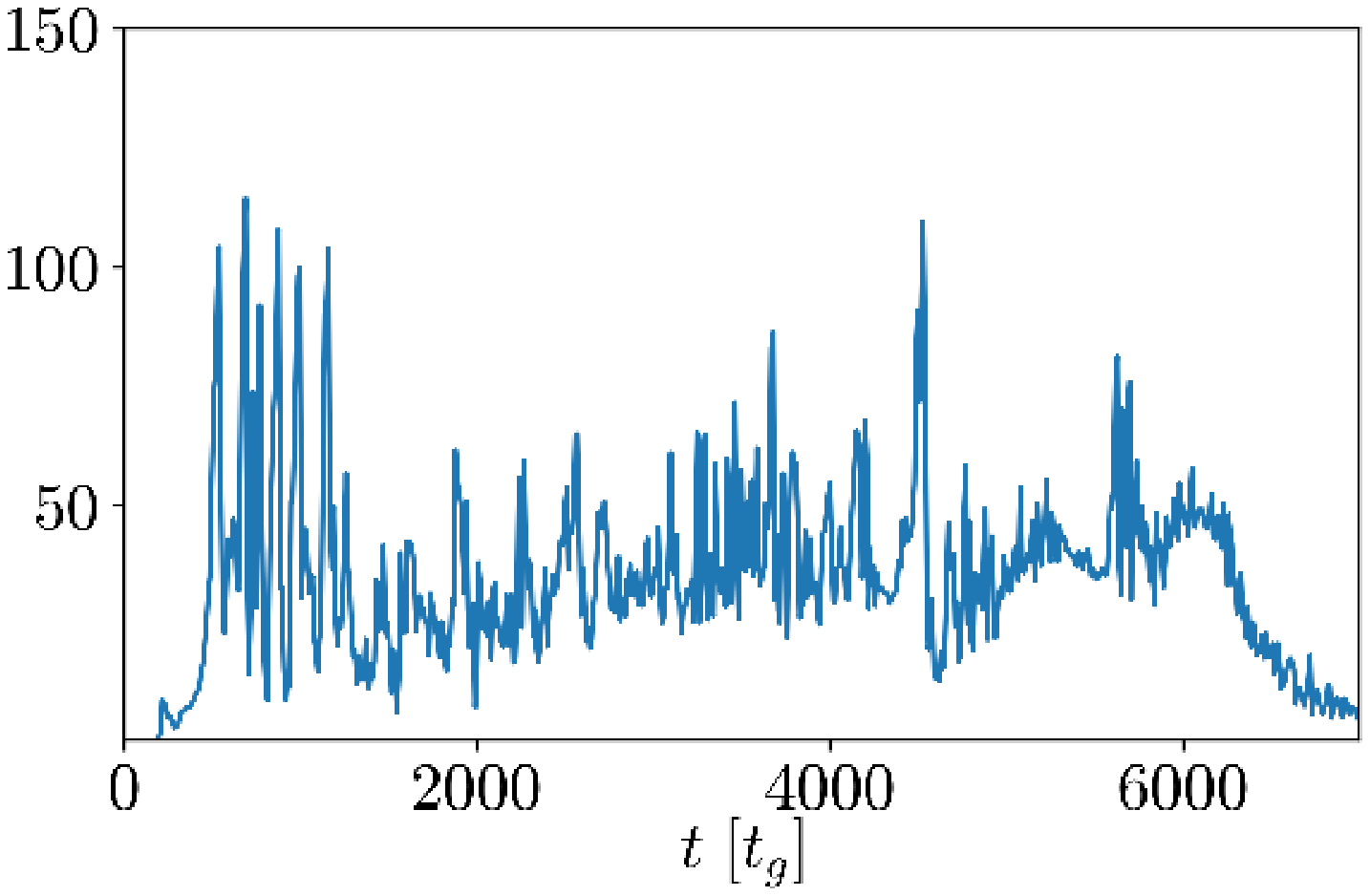}
    \quad
    \includegraphics[width=0.23\textwidth]{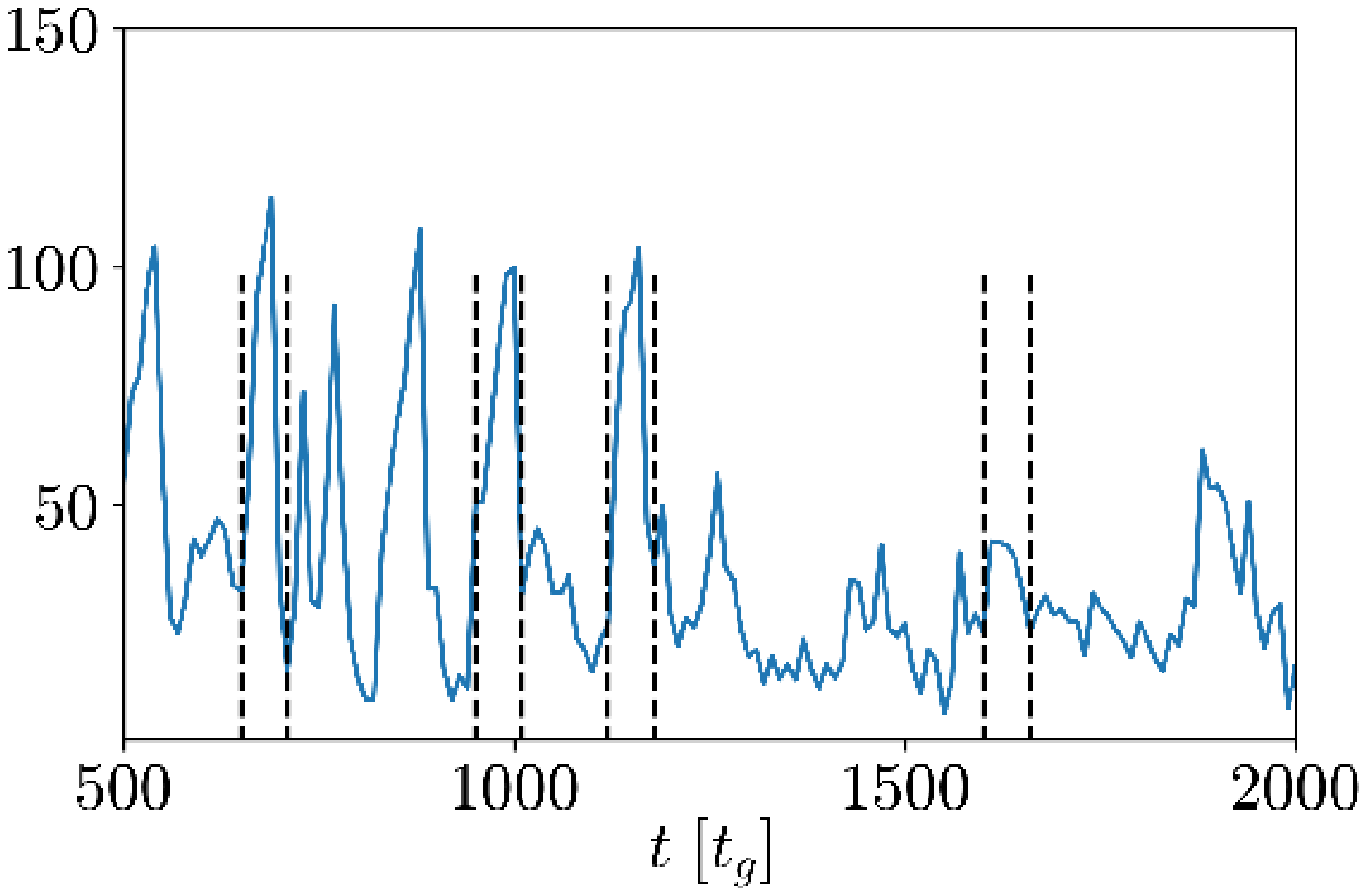}
    \quad
    \includegraphics[width=0.23\textwidth]{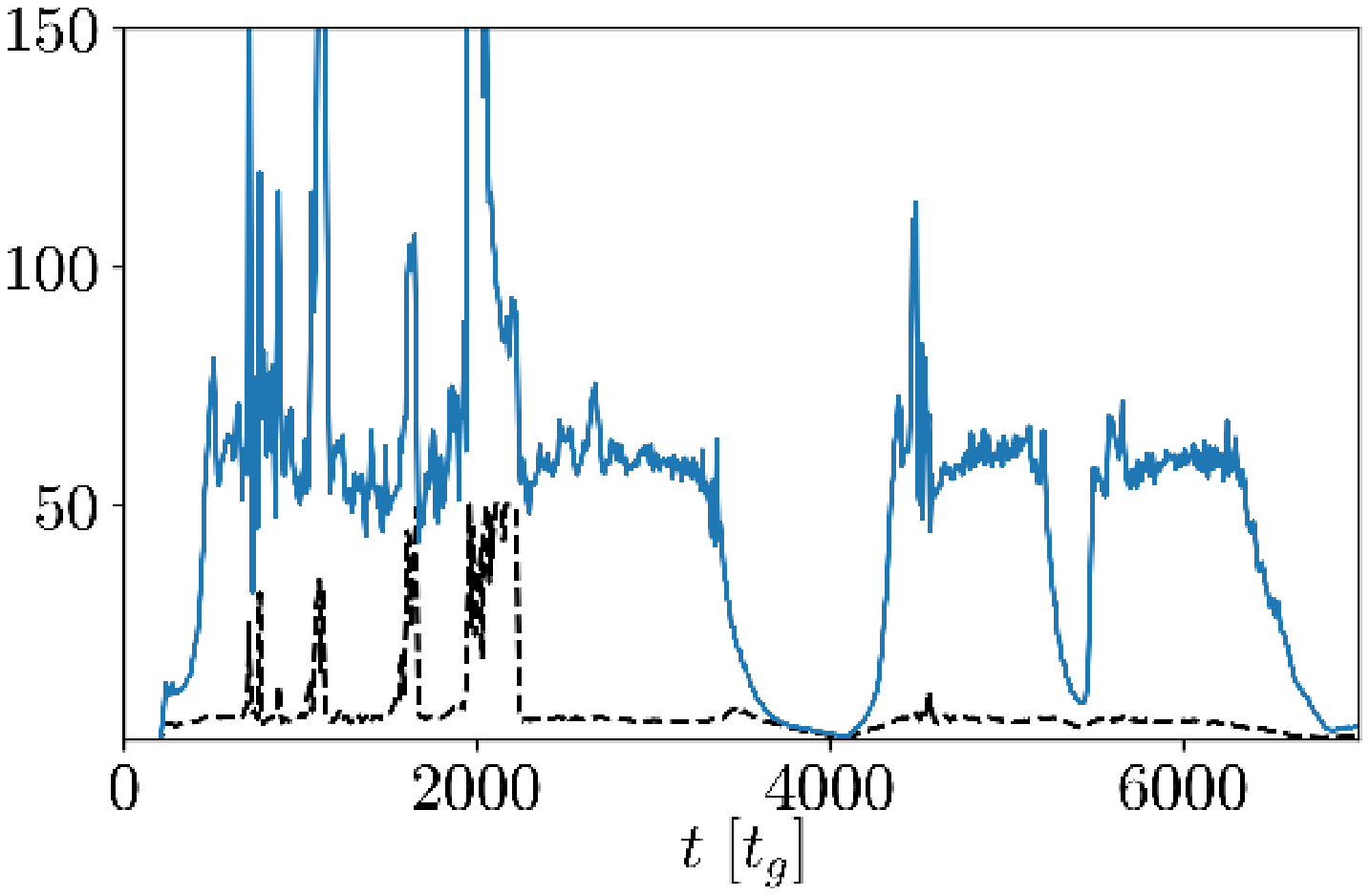}
    \quad
    \includegraphics[width=0.23\textwidth]{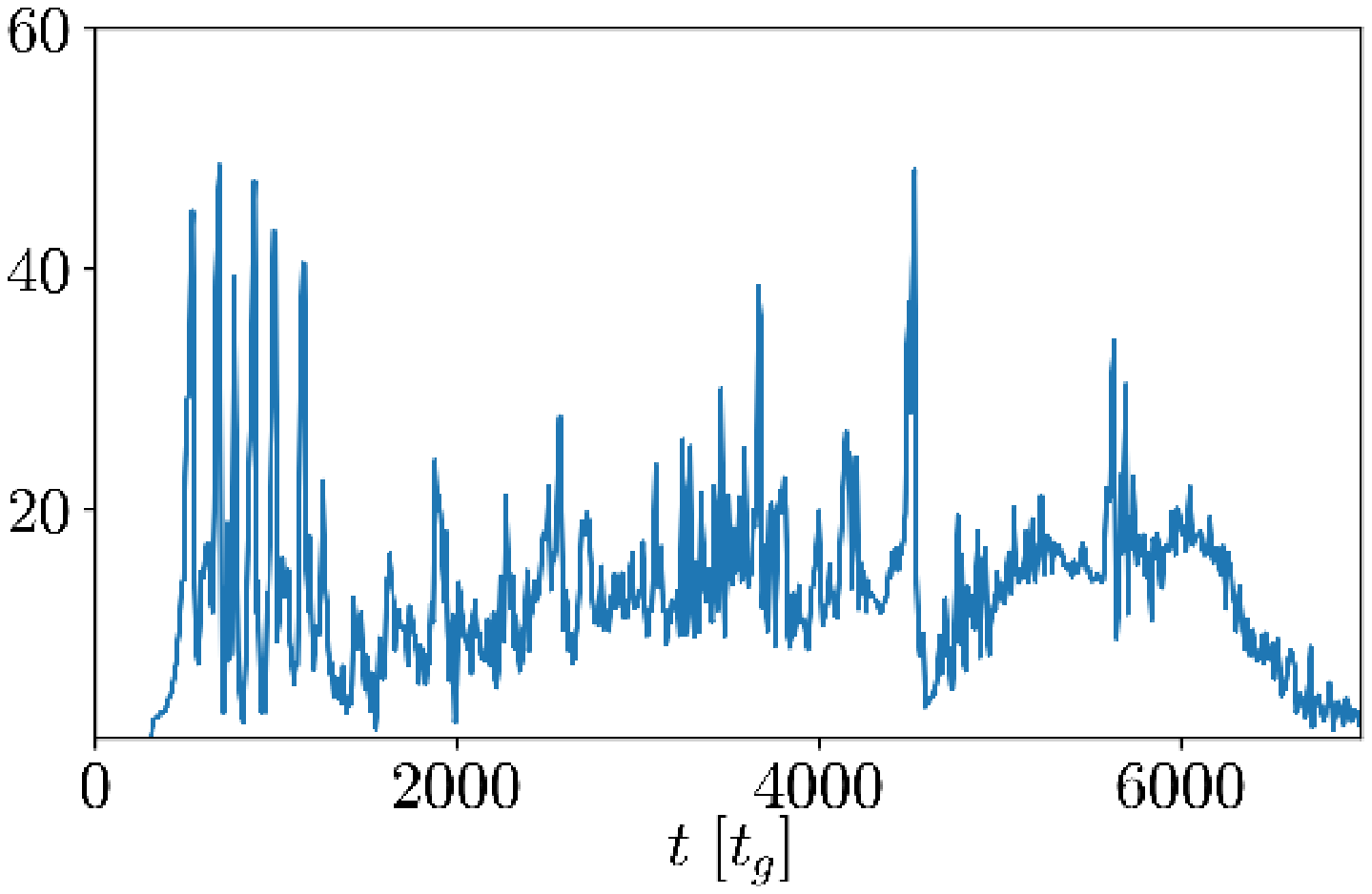}
    \\
    \includegraphics[width=0.23\textwidth]{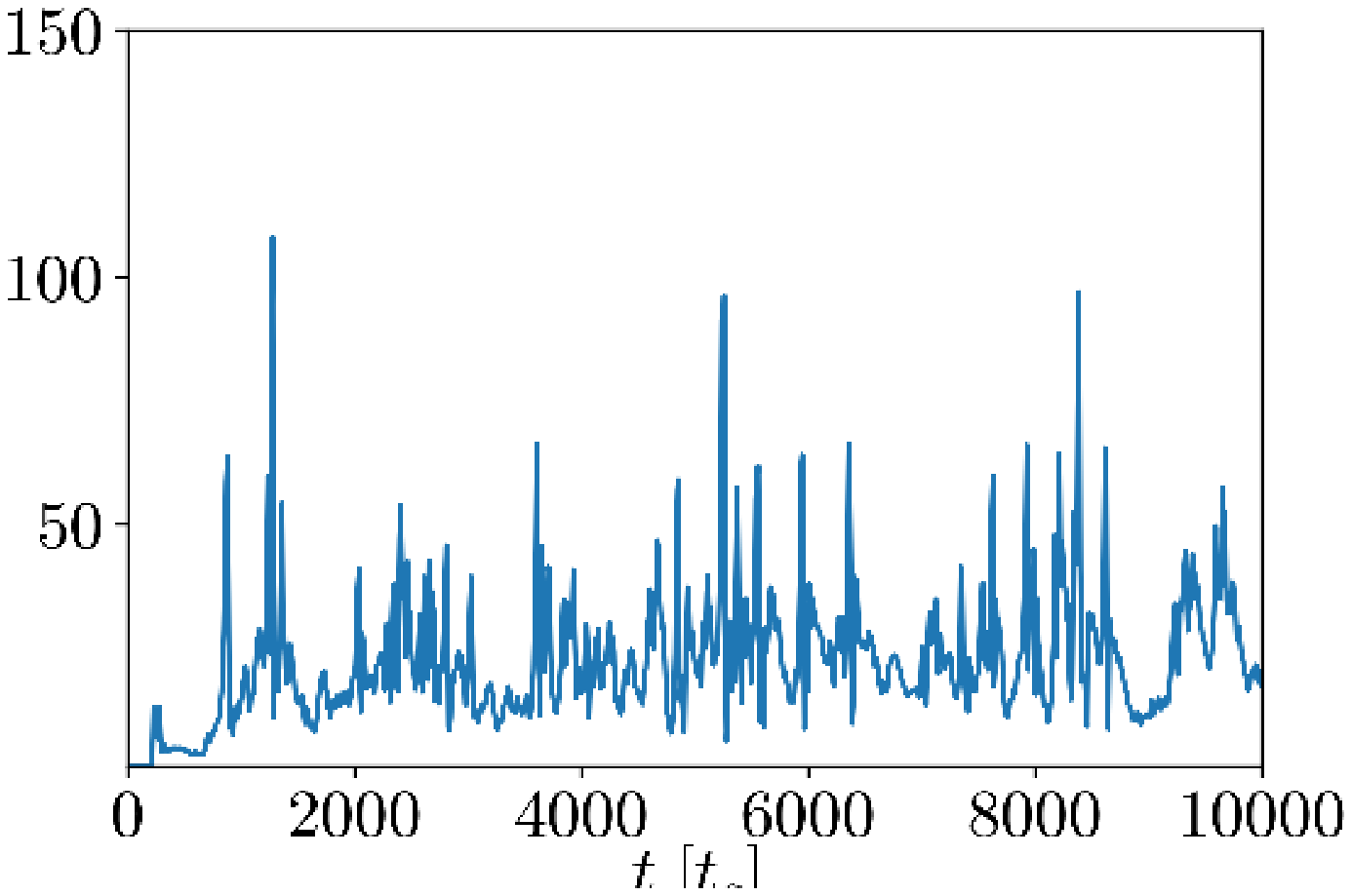}
    \quad
    \includegraphics[width=0.23\textwidth]{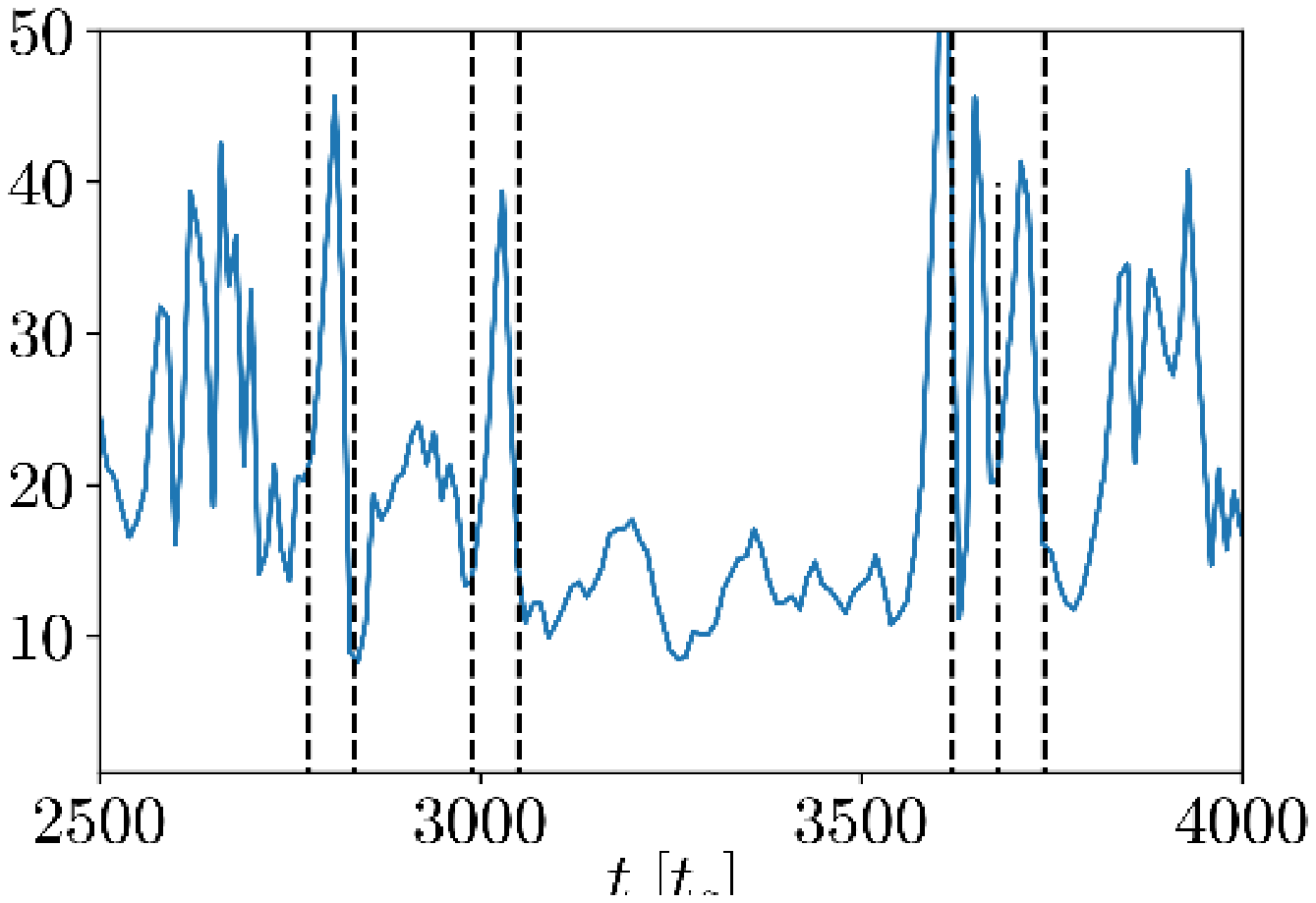}
    \quad
    \includegraphics[width=0.23\textwidth]{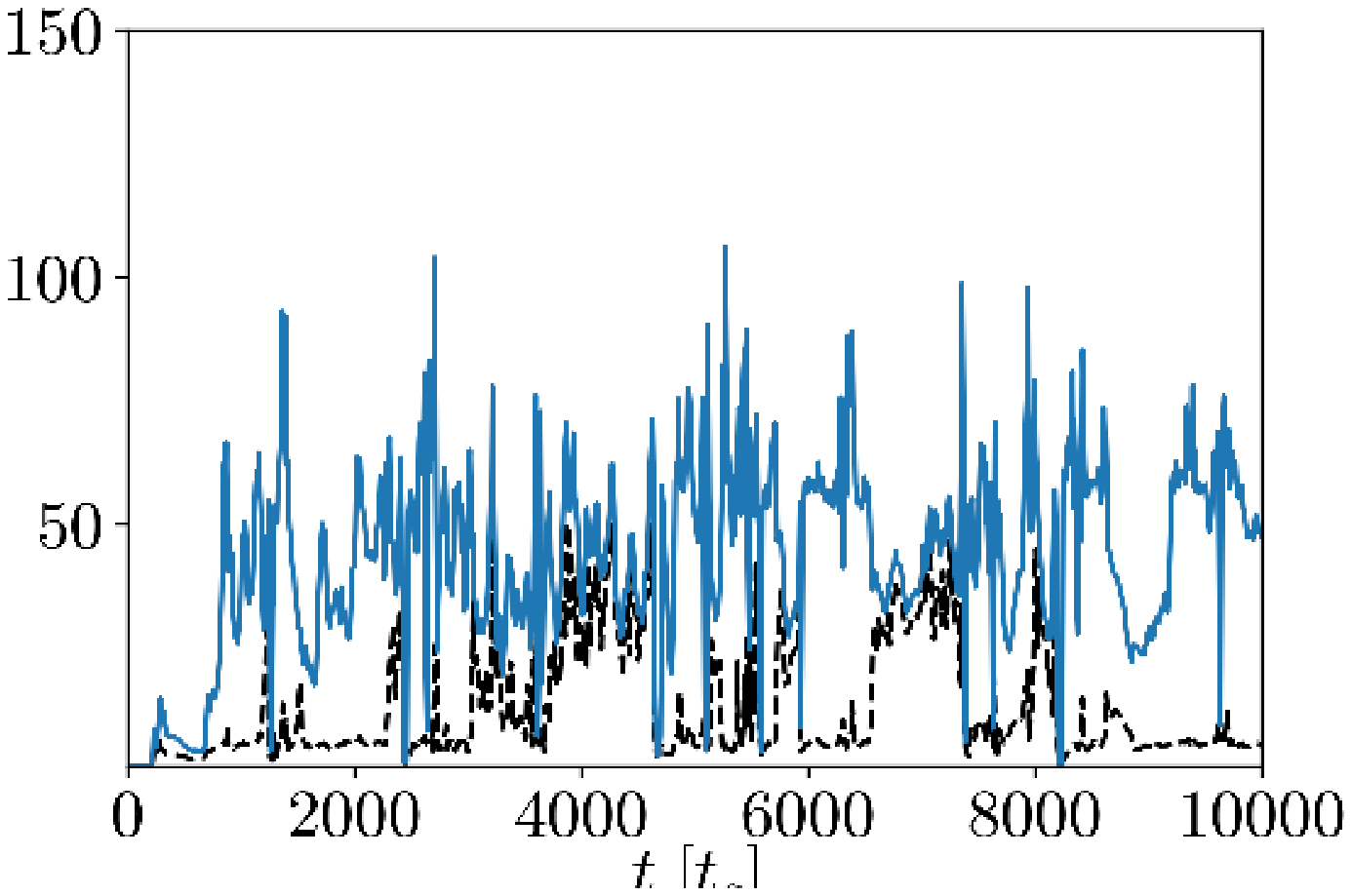}
    \quad
    \includegraphics[width=0.23\textwidth]{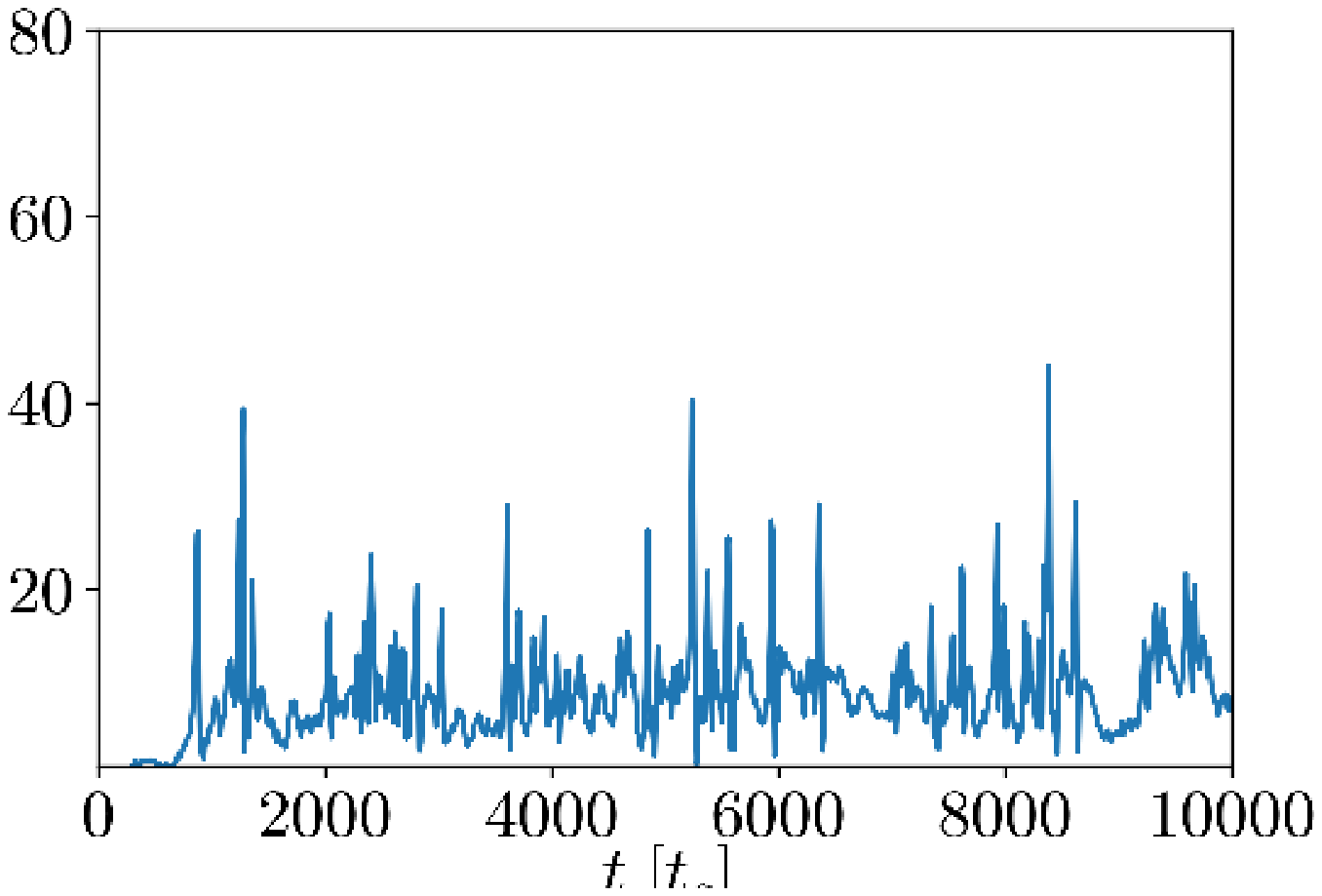}
    \\
    \includegraphics[width=0.23\textwidth]{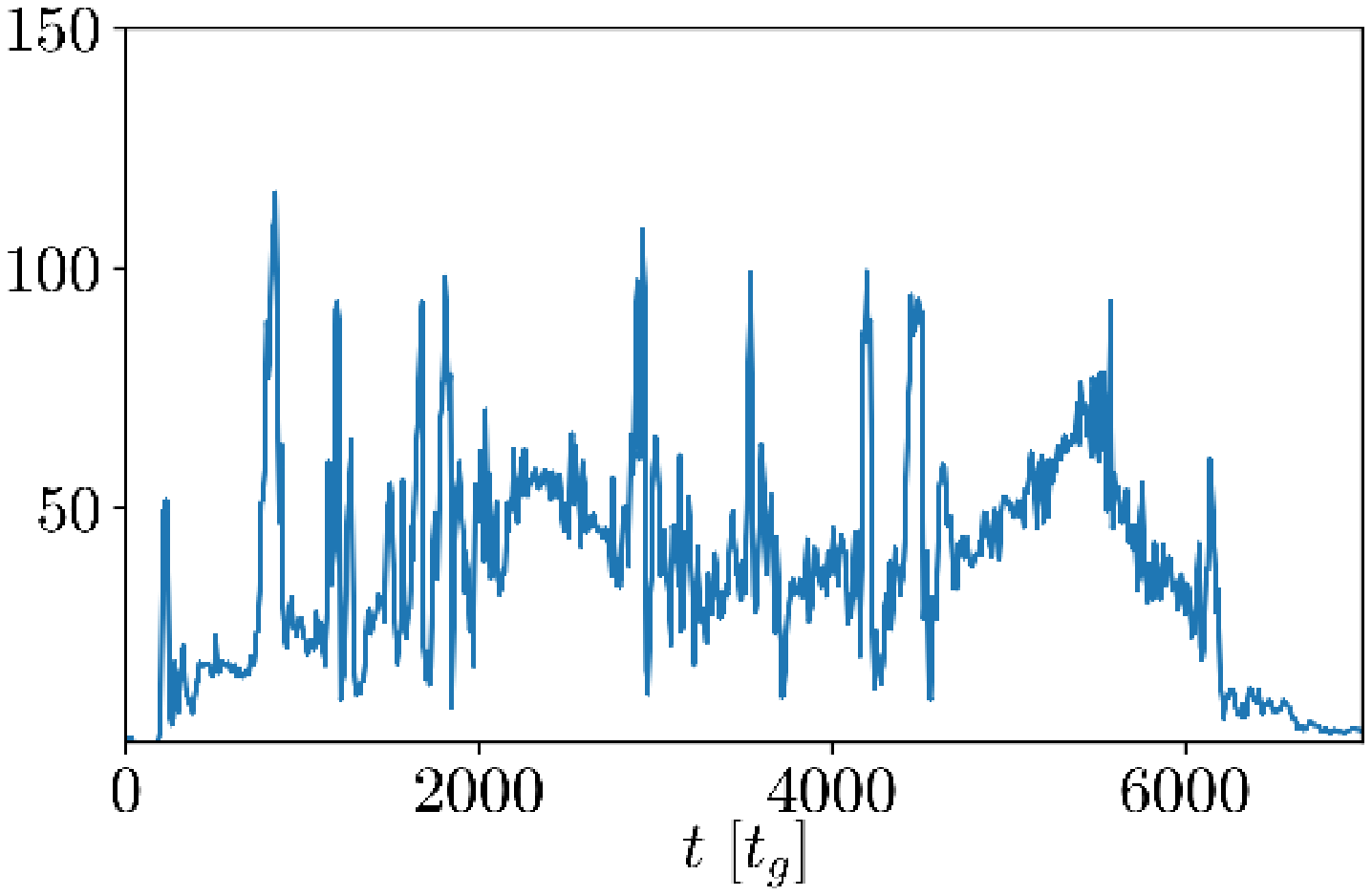}
    \quad
    \includegraphics[width=0.23\textwidth]{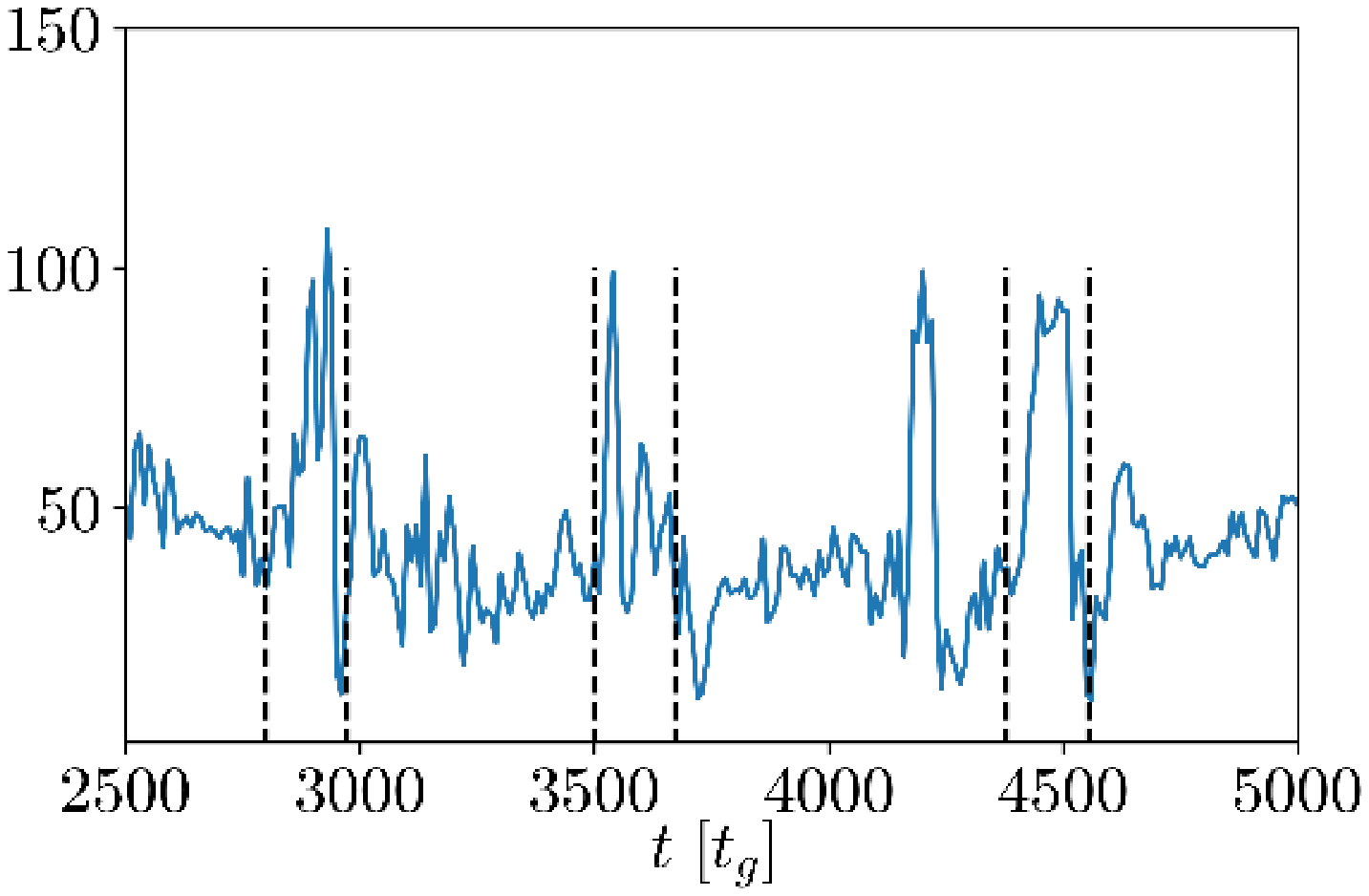}
    \quad
    \includegraphics[width=0.23\textwidth]{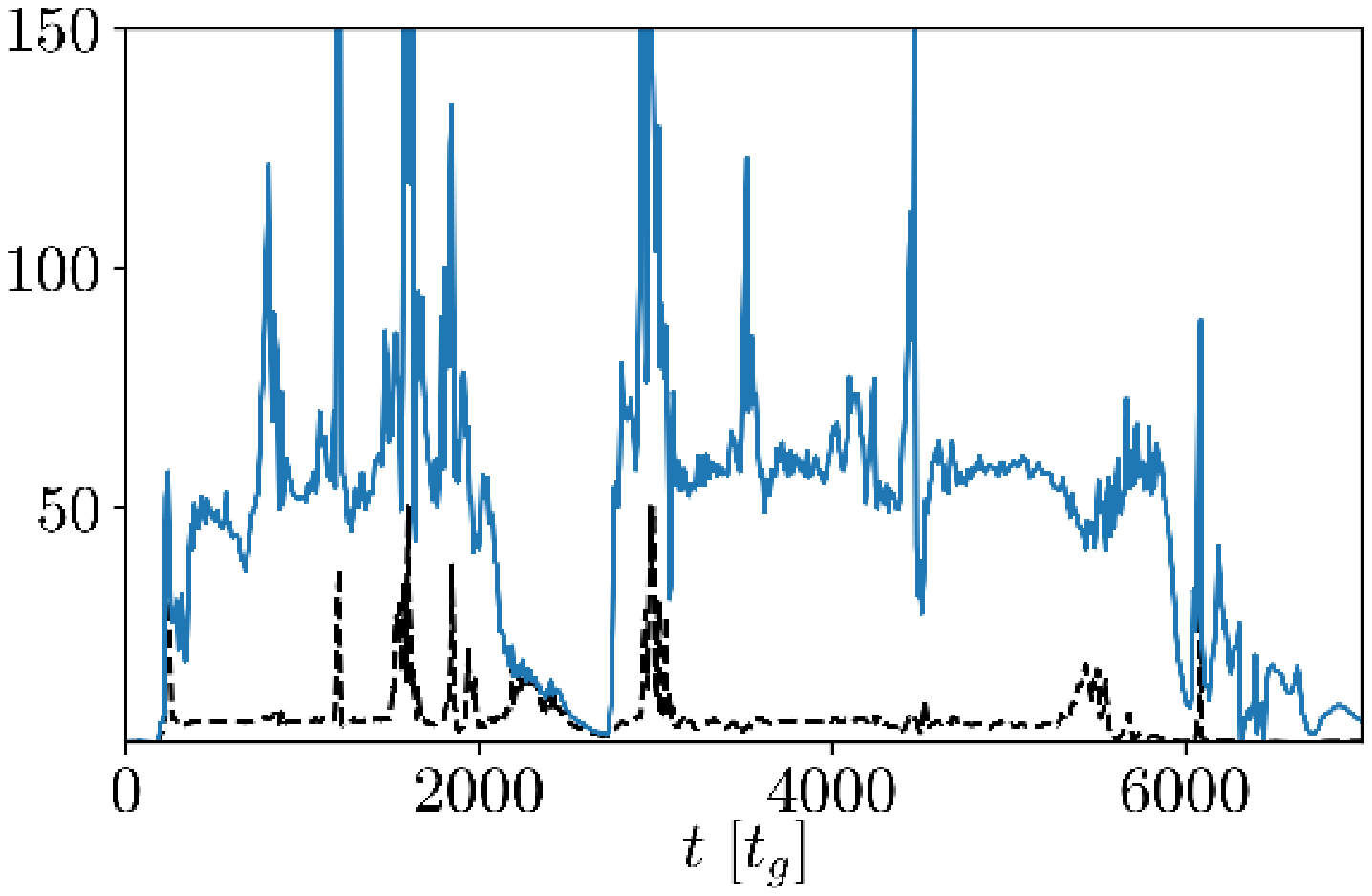}
    \quad
    \includegraphics[width=0.23\textwidth]{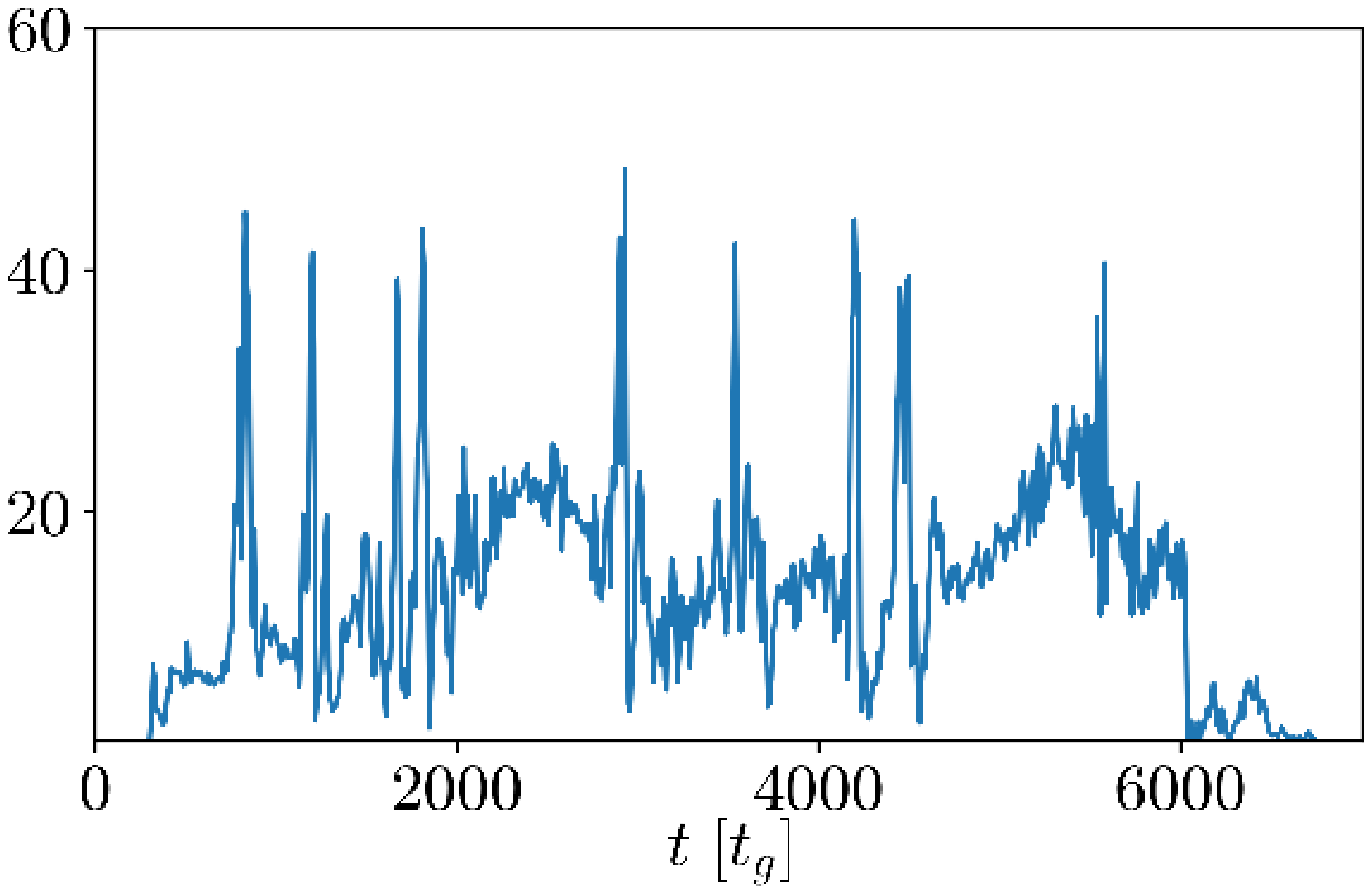}
    \\
    \includegraphics[width=0.23\textwidth]{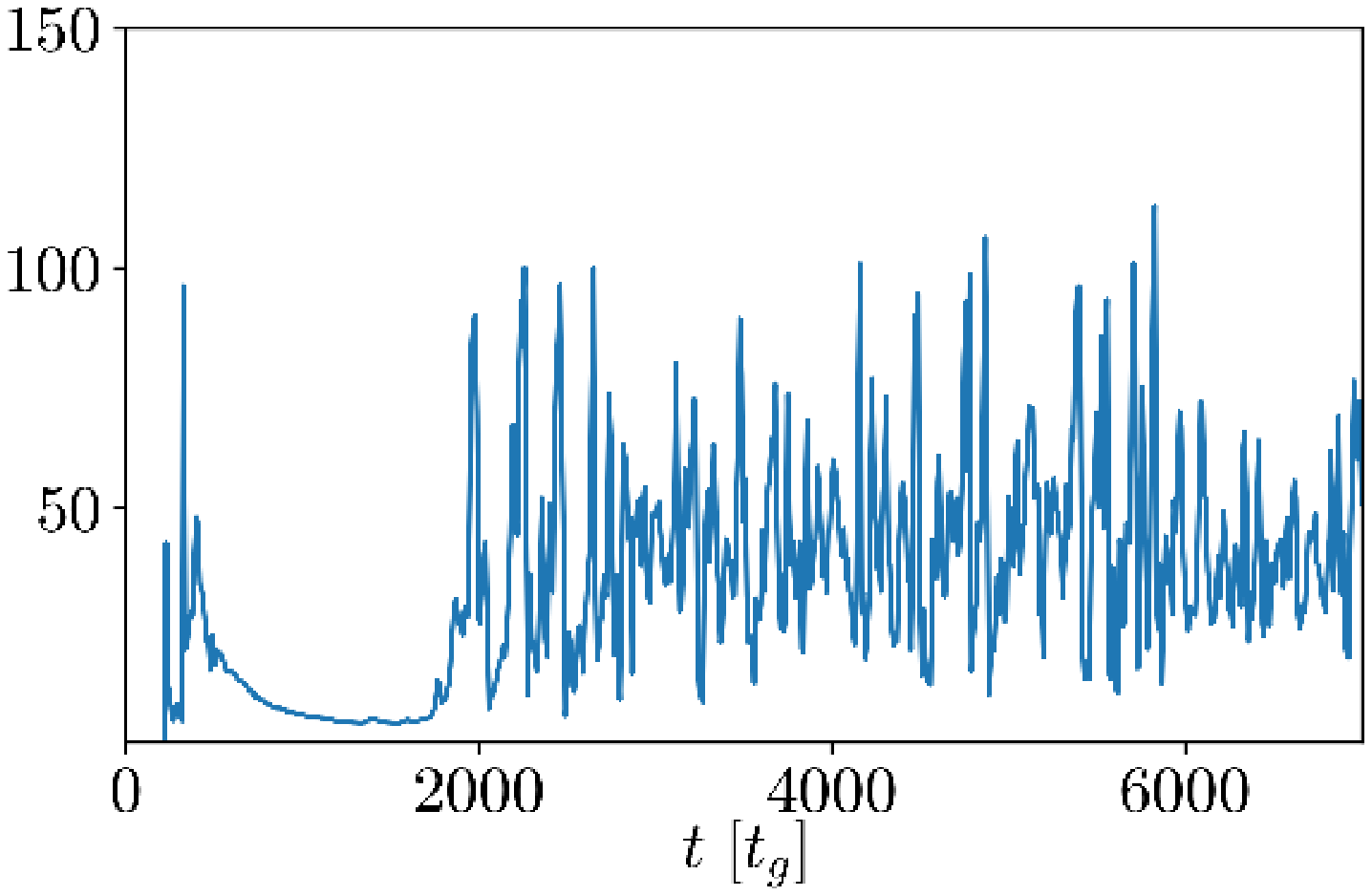}
    \quad
    \includegraphics[width=0.23\textwidth]{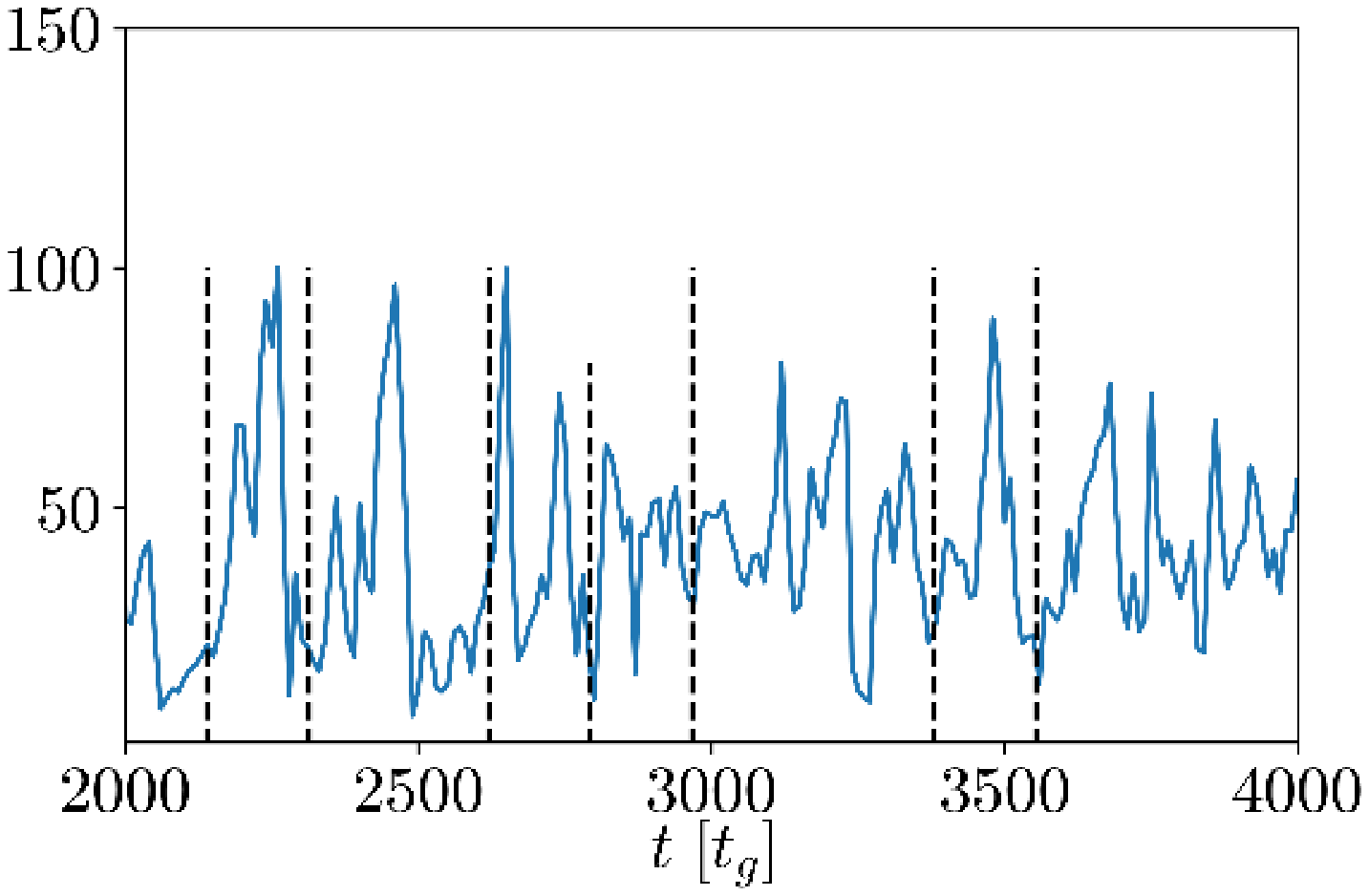}
    \quad
    \includegraphics[width=0.23\textwidth]{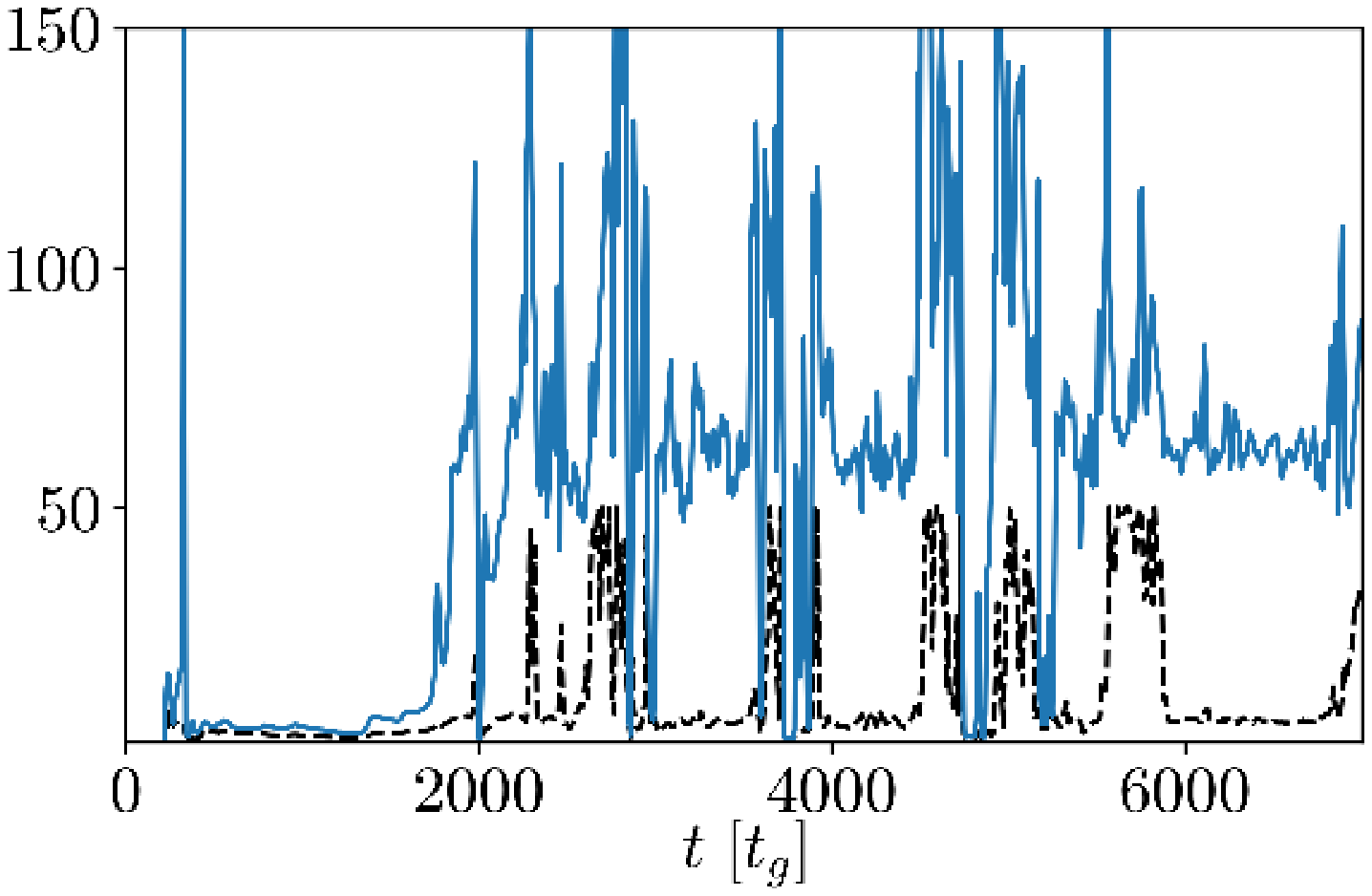}
    \quad
    \includegraphics[width=0.23\textwidth]{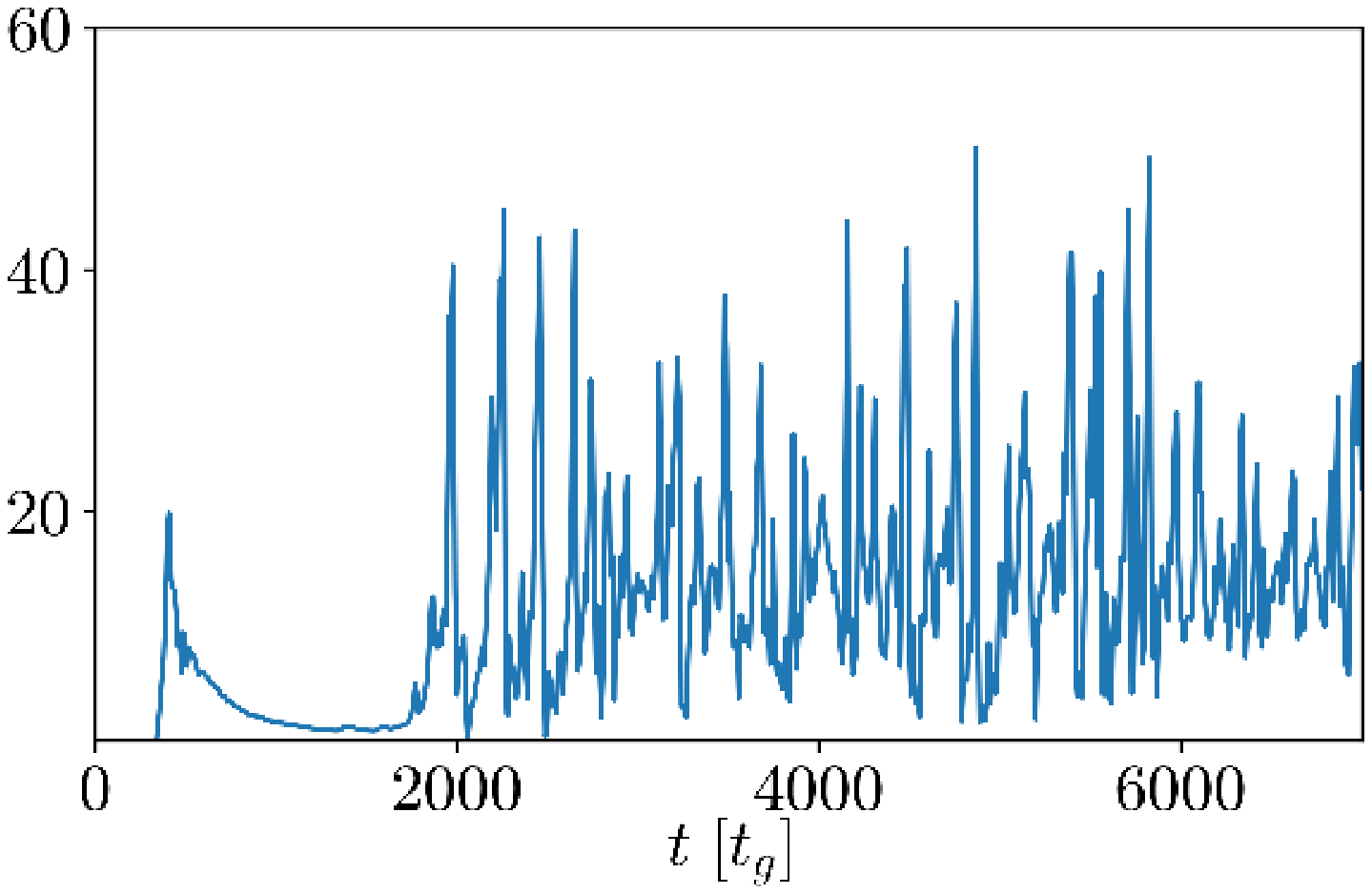}   
    \\
    \includegraphics[width=0.23\textwidth]{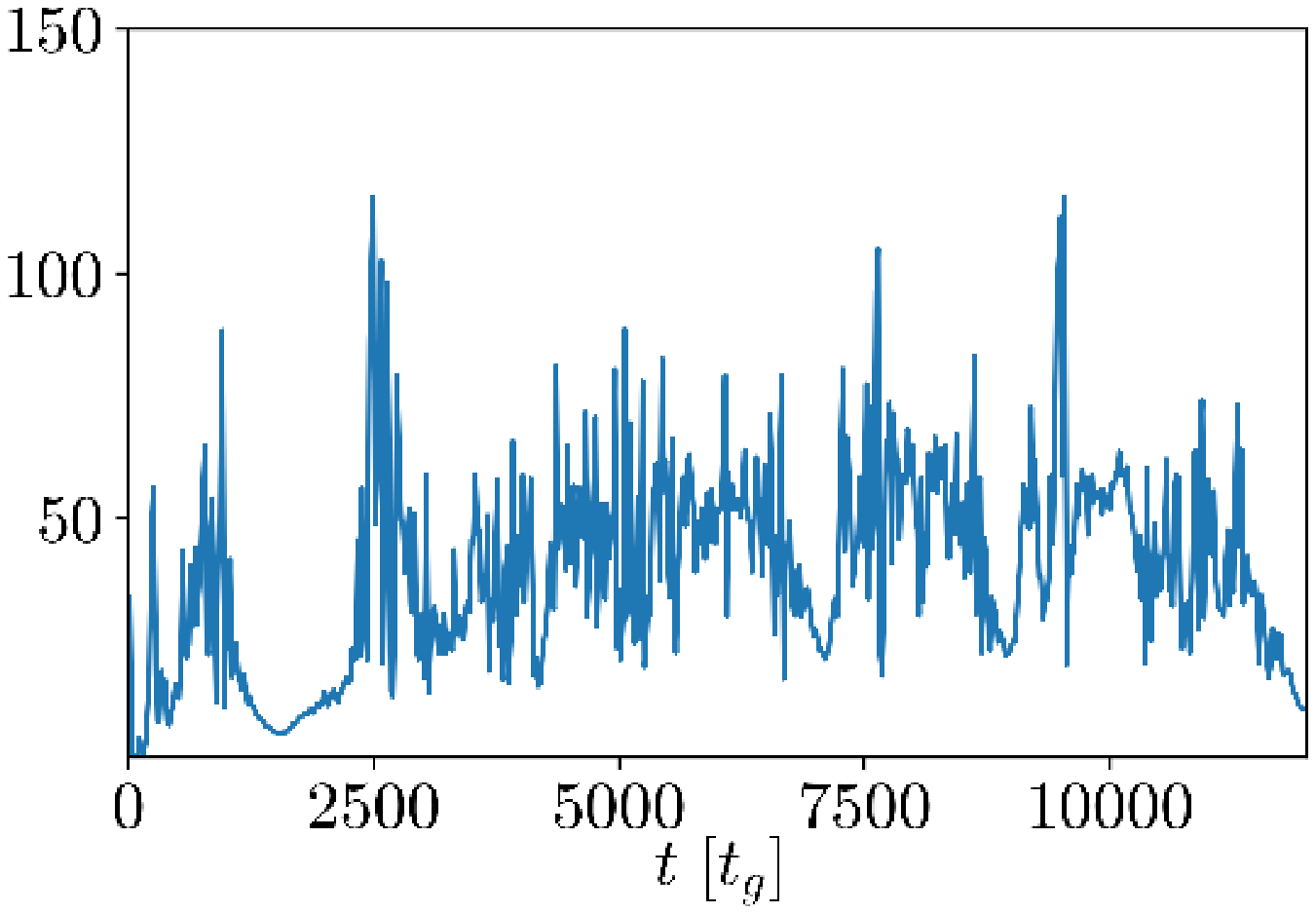}
    \quad
    \includegraphics[width=0.23\textwidth]{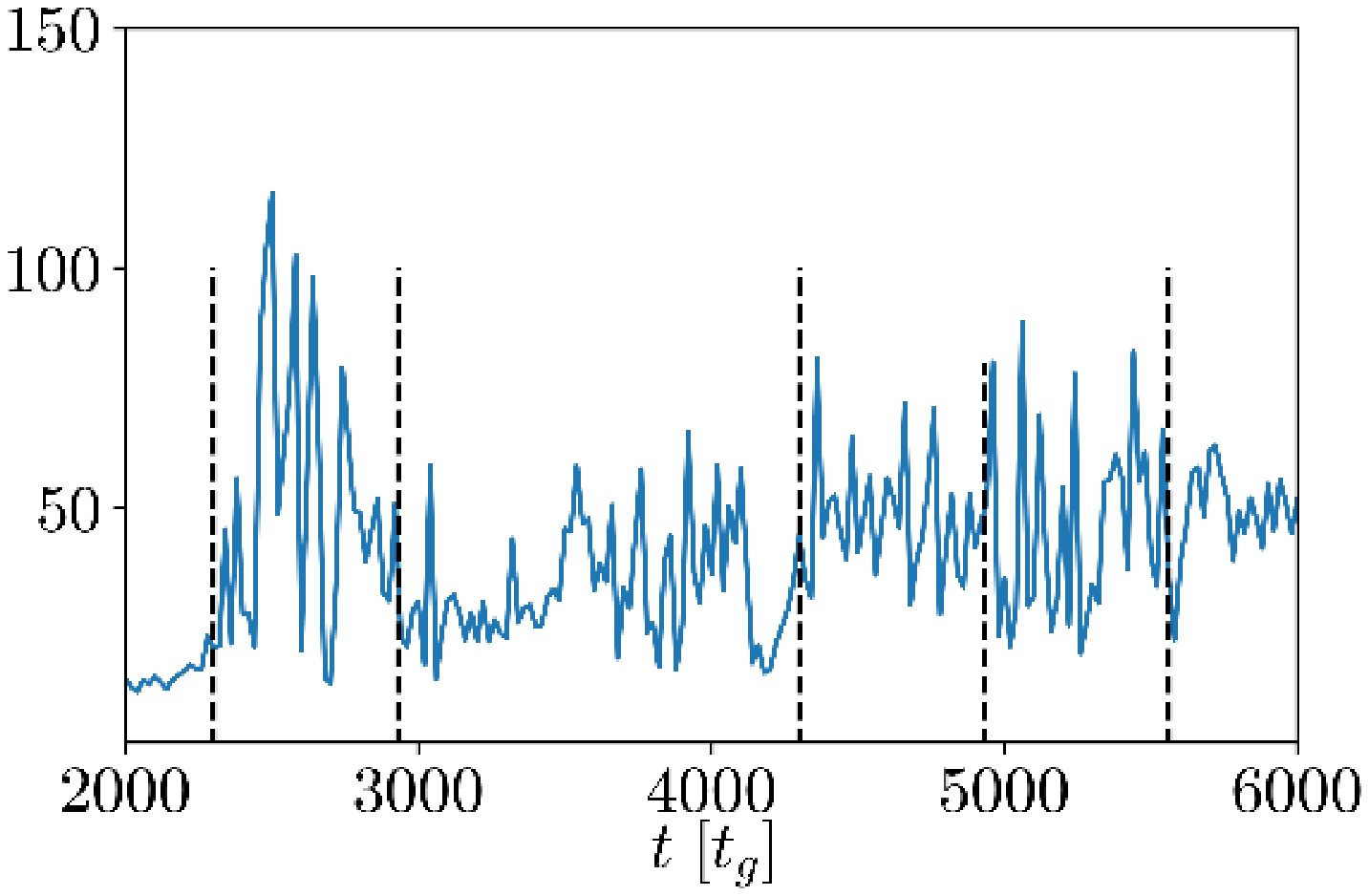}
    \quad
    \includegraphics[width=0.23\textwidth]{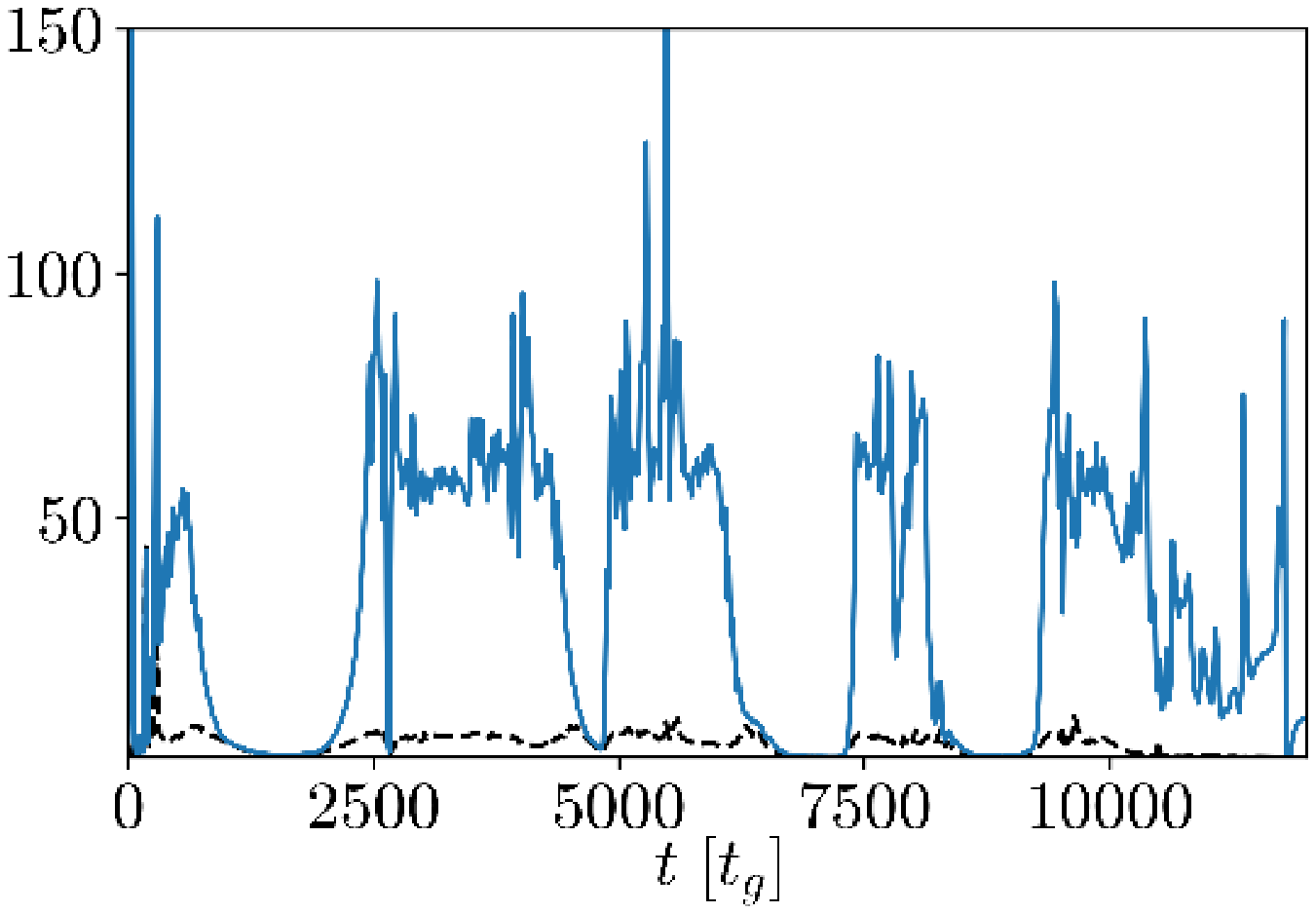}
    \quad
    \includegraphics[width=0.23\textwidth]{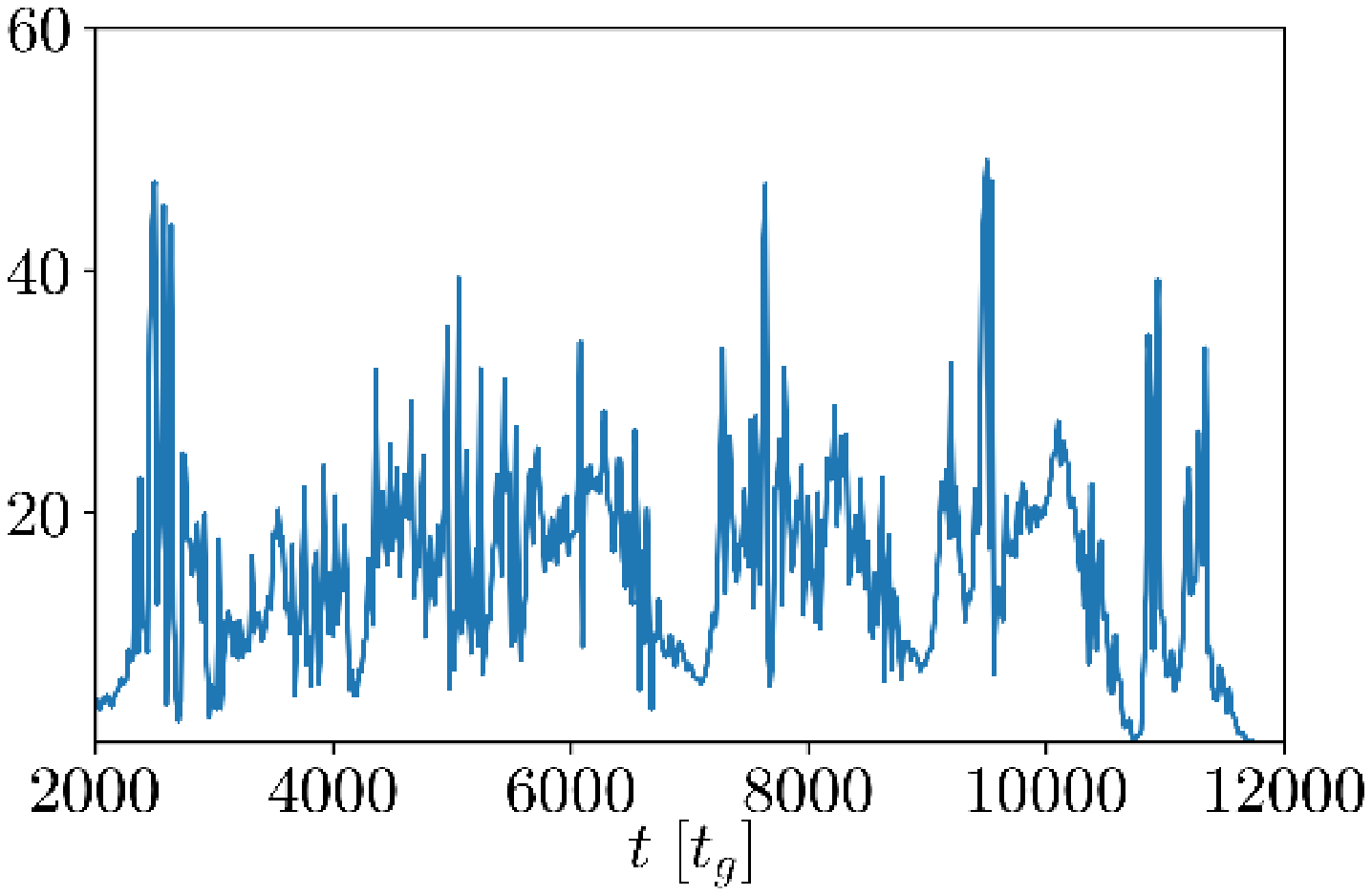}
    \caption{The results from top to bottom for the \textit{LD-Magn, LD-Therm, MD-Magn, MD-Therm, HD-Magn} models. \textbf{1st column:} The energetic parameter $\mu$ for the interior $p_1$ point of reference. \textbf{2nd column:} a zoom in time of the previous column diagrams. The dashed lines represent the characteristic time scale of the MRI and shows an accurate accordance in the first 4 models, while shows no obvious correlation at the \textit{HD-Magn} model. \textbf{3rd column:} the energetic parameter close to the jet boundary $p_2$. The dotted lines of the later diagrams stands for the lorentz factor illustrating the partial acceleration at the $z\sim200r_g$. \textbf{4th column:} The magnetization parameter $\sigma$ at the $p_1$ point illustrating the magnetic dominance of the jet.}
    \label{fig:models_resutls}
\end{figure*}

\subsection{Time variability of the jet}

\indent The high inhomogeneity of the outflow is present in all the quantities but its implication is better illustrated in the $\mu$ and $\sigma$. We use the former quantity to perform the time variability analysis. The point of the outflow is chosen to be far enough so that the space-time is flat and the quantity is close to its special relativistic interpretation. It must be close enough so that the logarithmic scale of our grid is dense enough to describe the outflow reliably. Moreover, the higher acceleration close to the jet boundary produces a less sharp variability than in the inner region making the analysis more complicated. As a consequence we considered two points of reference, an inner one $p_1: (x,z)=(11.5,200)r_g$ to perform the time variability and an outer one $p_2: (x,z)=(40,197)r_g$, to identify acceleration (see Figure~\ref{fig:models_resutls}). Since our outflow is not in a steady state, the synchronization of the space-time points sets an extra concern for the points to be compared \citep{Fontetal1999, Garofaloetal2010}. Nevertheless, the large distance of the points selected minimizes the effects of this implication.

\indent The intense variability of the outflow is evident in all models. The second column shows a zoom in time for the $p_1$ point and the relevant interval corresponding to the characteristic time of the maximum growth as calculated by Eq.~\ref{eq:growthrate}. The inspection of the diagrams shows that the quality of fitting the outflow variability with the $T_{\rm MRI}$ decreases as the distance of the initial torus increases. Thus, in the \textit{LD} models the correspondence is accurate, while in the \textit{MD} still in good agreement. In the \textit{HD} models it is almost absent and the outflow variability follows a much shorter time scale with many peaks included inside the MRI interval. In the same diagrams we also notice the low acceleration of the outflow, as it is expected by the special relativistic jet theory. The acceleration seems a little more enhanced at the boundary surface, probably because of the interaction with the exterior environment, but still insufficient. The longer time scale variability observed at the exterior point (see \textit{HD} in Fig. \ref{fig:models_resutls}) corresponds to narrower jet cross sections at the specific time intervals. The strong magnetization of the outflow is presented at the last column of the Fig.\ref{fig:models_resutls}, where values of $\sigma\sim 10-50$ are obtained for all the models.

\begin{figure}
    \centering
     \includegraphics[width=0.45\textwidth]{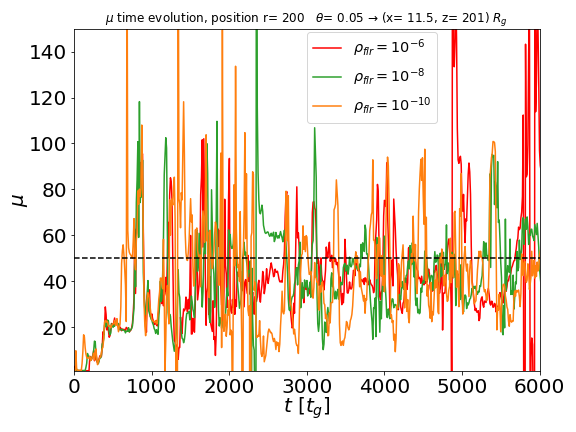}
     \includegraphics[width=0.45\textwidth]{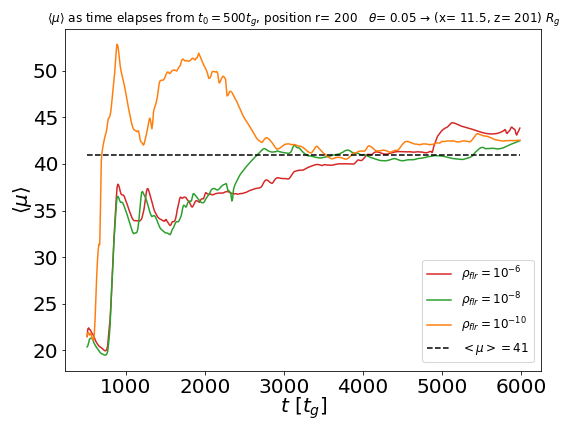}
     \caption{The time variability of the jet energy $\mu$ in the \textit{MMD-Magn} model, for three different values of the density floor. The left panel shows evolution of $\mu$ inside the jet, as taken at the distance of 200 $R_{\rm g}$. The right panel shows the value as averaged from $t_{0}=500$ until $t_{f}=t$. } 
     \label{fig:mu_floor}
\end{figure}

\indent Finally, we tested whether the results depend on the value of the adopted density and internal energy floor, and in particular, what is the floor influence on the maximum resulting energy parameter, $\mu$, as given by Eq. \ref{eq:mu}. In the Figure \ref{fig:mu_floor} we present an example run for the MD-Magn model with time variability of the $\mu$ value, for three different levels of the density floor.  
In addition, we show the time average, which is overpassing the fast variability of the $\mu$. The average is taken from $t_{0}=500$ until $t_{f}= 1000$, 2000, etc., giving the function of $<\mu>(t)$. One can see that the average value of $\mu$ saturates always at the same level, and the density floor, contrary to the force free floor $\beta_{\rm min}$, doesn't play an important role. This was expected since we have a very magnetized jet outflow ($\beta \sim 10^{-4}$) and the mass injected when the floor is reached is very small. Moreover, we should notice that the physical significance of $\mu$, in special relativistic theory, is the ratio of the total energy flux to the mass flux, so the mass normalization is in fact included in its definition.

\subsection{Jet power}

\begin{figure}
    \centering
     \includegraphics[width=0.45\textwidth]{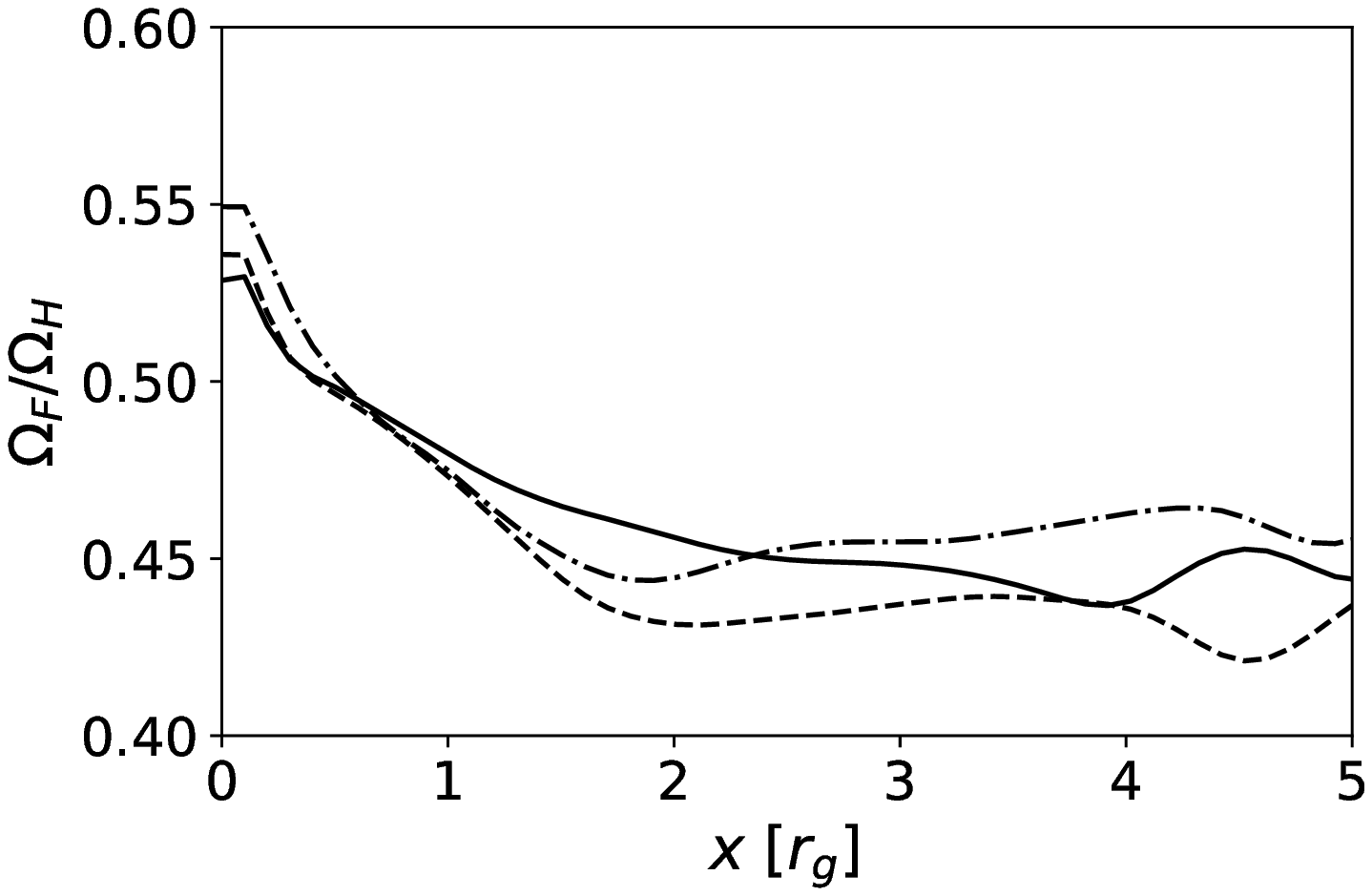}
     \caption{The ratio $\Omega_F / \Omega_H $ at $z=7 r_g$ and $t=3000t_g$ for the \textit{Magn} models as a function of the $x=rcos\theta$ coordinate. The values of $0.45-0.55$ across the funnel are in accordance with the BZ operation.} 
     \label{fig:BZ_snapshot}
\end{figure}

\indent The last aspect we investigate is the consistency of our simulation with the Blandford-Znajek mechanism. For that purpose we use the Faraday tensor $F_{\mu\nu}$ to calculate the rotational velocity of the magnetic field $\Omega_F = F_{t\theta}/F_{\theta\phi}$ and compare it with the angular frequency of the black hole $\Omega_{BH}=\left(a / 2\right)\left(1+\sqrt{1-a^2}\right)$. Figure~\ref{fig:BZ_snapshot} illustrates their mutual ratio at a distance $z=7 r_g$ from the black hole. Following the calculations of \citet{Yang_2015}, the outflow inside the jet funnel is close to the force free limit and the poloidal magnetic field keeps a parabolic configuration. Following their relaxation solution in case of the Blandford-Znajek operation we expect inside the funnel $\Omega_F / \Omega_H \sim 0.4-0.5$ in the case of slowly rotating black holes $a \sim 0.1$. The \citet{Komissarov2001} calculations for a monopole reveal a similar range of values $\Omega_F / \Omega_H = 0.5-0.55$ inside the funnel for a highly rotating Black Hole ($a \sim 0.9$), like the one assumed in our case. As a result the range of the ratio should not be far from these values. In our case inside the funnel the range $\Omega_F / \Omega_H \sim 0.53-0.4$, except some small region close to the outer boundary. Therefore we conclude that the BZ mechanism indeed operates, while differences from the values mentioned above are ascribed to the deviation of our solution from the steady state and the combined implications of the high black hole spin and our field geometry. A more extensive parametric study of the effects of these quantities, which are related to the past history of the pre-merger system will be postponed to the future work. We note here only that the value of black hole spin derived from the NS-NS merger simulations might be treated as the upper limit, since these simulations typically do not extend after the hypermassive neutron star is collapsed, while it may still loose a significant part of its angular momentum \citep{2014MNRAS.443.3134P, 2014MNRAS.441.3444M}.

\section{Discussion and conclusions}

\indent The extended magnetic field configuration adopted in our model is very efficient in the formation of a magnetic barrier and the resulting launch of a highly magnetized and low baryon loaded jet. The required accumulated magnetic flux is quickly piled up at the initial steps of the integration even for the \textit{Therm} model, where the initial $\beta$-parameter has rather high values. The precise magnetization of the initial torus is not crucial for the emerging Poynting outflow and although it affects the time needed for the magnetic barrier to accumulate the necessary flux, the final outcome shows only a slight preference for the \textit{LD-Magn} models. The operation of the central engine lasts until the pressure of the accreting gas becomes much lower than the pressure of the magnetic field threading the BH. From that point and beyond, a magnetically arrested disk form (MAD, \cite{Tchekhovskoy_Giannios2015}). The accurate description of this stage should be studied by 3D simulations, but a qualitative indication can be obtained also in 2D since the \textit{Therm} models reach the arrested stage in longer times. Further analysis of this problem is the subject of our future work.

\indent Another difficulty we encountered when an even more enhanced magnetic content of the initial torus, was the formation of an extensive barrier that disrupts the accreting system. We bypass this problem by assuming an ISM of higher density.
The implications of the ISM initial state is an interesting topic by itself and needs further clarification, but such a parametric study falls outside the scope of the current work. Of course this does not exclude the possibility of a different magnetic topology and a more highly magnetized torus, if a portion of the exterior field has been accreted, or expelled, at the phase prior to the BH-torus formation. Such a scenario naturally arises in the NSNS type of progenitors (see \citet{Ruiz2016}, where the collapsed HNMS carries the magnetic flux of its predecessor). 

\indent In comparison with the 3-dimensional simulations of the GRB jets, performed recently, e.g. by \citet{2018arXiv180905099K}, our axisymmetric simulations produce a similar internal structure of the jets, although these studies are not focused on the time variability analysis and show only time averaged profiles of the terminal Lorentz factor. 
We notice, that the full description of physical processes within the jets with an arbitrary magnitude of all components of magnetic field vector is possible only in 3-dimensional setup. Nevertheless, if the component of $B_{\phi}$ is smaller than two other, the jet can be successfully launched and sustained for a long time of the simulation, as shown by our work. Although the 2D simulation by definition suppress the toroidal wave numbers, our results do not show any MRI diffussion. 

\indent The 3D description of the MRI differs indeed from the 2D case in two ways. The toroidal wave numbers and disturbances are suppressed by axisymmetry. This is not of primary importance to us since we do not have the $B_{\phi}$ component initially. In fact, the wavenumbers in the $\phi$ direction might result in a  deeper or shallower variability in 3D simulation, but they would not lead to a totally different pattern. We conclude that the application of our method in 3D simulations can provide some further insight into the problem. Currently this is beyond the scope of our work but we postpone the computations to the future investigations. 

\indent The second important problem is the decay of the MRI instability because of the Cowling's anti-dynamo theorem. The time of the decay is however affected by the resolution (see Figures 9 and 10 in the article by \citet{2008ApJS..174..145G}). For a really high resolution (as the one we adopt) the decay is minimal and this is why we chose it. Moreover, if the instability was decaying, or some other parasitic modes were dominating, it would have an impact on the time variability of the outflow, which we do not observe.

\indent The analysis of the outflow inhomogeneity was performed in terms of the energetic parameter $\mu$, being the proxy of the Lorentz factor itself. This specific quantity provides a significant advantage if the emerging outflow is consistent with the special relativistic theory of jets, like in our case. The special relativistic prediction that the Poynting flux is converted to bulk kinetic in a larger spatial scale challenges in general the simulations studying the jet launching process. Although the accurate final outflow state must be still obtained by a global or synthetic simulation able to cover the whole spatial regime, significant results can be obtained. In such a case the energetic parameter considered at large distance from the gravitational sources, i.e. where space-time curvature is negligible, comes in handy.      

\indent In our model the energetic parameter $\mu$ provides an accurate description of the inhomogeneity and enables the study of the MRI implication. Resolving the MRI with a minimum of $10$ cells per wavelength, the results revealed an accurate correspondence for the shortest assumed torus distance, independently of the magnitude of the magnetic field. This is consistent with the theoretical calculations since the maximum growth of the instability does not depend on the magnetic field magnitude (Eq.~\ref{eq:growthrate}). The MRI is also in accordance with the variability of the \textit{MD} models although the inspection of the results reveals that inside a characteristic time scale there can exist double or triple peaks. On the contrary the correlation is ambiguous in the \textit{HD} models, where multiple peaks are found inside the characteristic time scale. We notice though that such a correlation can be associated with the extensive formations observed in the outer point of the $\mu$ variability. For our specific models the typical width of the spikes is better fitted by a $T\sim 70 t_g$ time scale which corresponds to the action of the instability at smaller radii in the accretion disk.

\indent Summarizing, our study associates the time variability of the jet with the dimension and the magnetic field topology of the torus that arises during the merging process. Assuming the formation of a $3 M_\odot$ BH, the \textit{LD} $\Delta t_{\rm blobs} \sim 8.9 \cdot 10^{-4} s$ and \textit{MD} $\Delta t_{\rm blobs} \sim 2.6 \cdot 10^{-3} s$ are both consistent with the reported range of the phenomenon variability, while the \textit{HD} models provide a large pattern of variability $\Delta t_{\rm blobs} \sim 10^{-2} s$ that it is not observed in general. 

\indent An extended debate occurs for the main process that drives the jet acceleration, i.e. the action of the Blandford-Znajek process versus the Blandford-Payne one. The consistency of $\Omega_F/\Omega_H$ (see Fig.\ref{fig:BZ_snapshot}) with the theoretical calculations establishes the operation of the B-Z process in the interior points of the outflow, while for the exterior points there is a strong indication that the BP must also contribute. In order to get some intuition of the relative power of the two mechanisms we calculated the time average of the magnetic parameter, obtaining $<\mu_1> \sim 41$ and $<\mu_2> \sim 64$. The magnitude of these values points to the conclusion of the comparable participation of both processes.
As we have shown above, the density floor does not affect these results in a great way, but the choice of $\beta_{\min}$ might affect the calculated value of $\mu$, especially for the strongest magnetized pulsations. As a result, the averaging procedure must be considered as an approximation under the specific code limitations.

\section*{Acknowledgments}
We thank Marek Sikora and Krzysztof Nalewajko for interesting discussions. 
We also thank the anonymous referee for constructive comments which helped to improve our manuscript.
This work was supported in part by the grants 
no. ~ DEC-2012/05/E/ST9/03914 and
 DEC-2016/23/B/ST9/03114, from the Polish National Science Center.
We also acknowledge support from the Interdisciplinary Center for Mathematical Modeling of the Warsaw University, through the computational grant Gb70-4.

\bibliographystyle{aasjournal}
\bibliography{kostas_grmhd.bib}

\begin{thebibliography}{}
\expandafter\ifx\csname natexlab\endcsname\relax\def\natexlab#1{#1}\fi
\providecommand{\url}[1]{\href{#1}{#1}}
\providecommand{\dodoi}[1]{doi:~\href{http://doi.org/#1}{\nolinkurl{#1}}}
\providecommand{\doeprint}[1]{\href{http://ascl.net/#1}{\nolinkurl{http://ascl.net/#1}}}
\providecommand{\doarXiv}[1]{\href{https://arxiv.org/abs/#1}{\nolinkurl{https://arxiv.org/abs/#1}}}

\bibitem[{Abbott {et~al.}(2017)Abbott, Abbott, Abbott, Acernese, Ackley, Adams,
  Adams, Addesso, Adhikari, Adya, Affeldt, Afrough, Agarwal, Agathos, Agatsuma,
  Aggarwal, Aguiar, Aiello, Ain, Ajith, Allen, Allen, Allocca, Aloy, Altin,
  Amato, Ananyeva, Anderson, Anderson, Angelova, Antier, Appert, Arai, Araya,
  Areeda, Arnaud, Arun, Ascenzi, Ashton, Ast, Aston, Astone, Atallah, Aufmuth,
  Aulbert, AultONeal, Austin, Avila-Alvarez, Babak, Bacon, Bader, Bae, Baker,
  Baldaccini, Ballardin, Ballmer, Banagiri, Barayoga, Barclay, Barish, Barker,
  Barkett, Barone, Barr, Barsotti, Barsuglia, Barta, Bartlett, Bartos, Bassiri,
  Basti, Batch, Bawaj, Bayley, Bazzan, B{\'e}csy, Beer, Bejger, Belahcene,
  Bell, Berger, Bergmann, Bero, Berry, Bersanetti, Bertolini, Betzwieser,
  Bhagwat, Bhandare, Bilenko, Billingsley, Billman, Birch, Birney, Birnholtz,
  Biscans, Biscoveanu, Bisht, Bitossi, Biwer, Bizouard, Blackburn, Blackman,
  Blair, Blair, Blair, Bloemen, Bock, Bode, Boer, Bogaert, Bohe, Bondu,
  Bonilla, Bonnand, Boom, Bork, Boschi, Bose, Bossie, Bouffanais, Bozzi,
  Bradaschia, Brady, Branchesi, Brau, Briant, Brillet, Brinkmann, Brisson,
  Brockill, Broida, Brooks, Brown, Brown, Brunett, Buchanan, Buikema, Bulik,
  Bulten, Buonanno, Buskulic, Buy, Byer, Cabero, Cadonati, Cagnoli, Cahillane,
  Calder{\'o}n~Bustillo, Callister, Calloni, Camp, Canepa, Canizares, Cannon,
  Cao, Cao, Capano, Capocasa, Carbognani, Caride, Carney, Casanueva~Diaz,
  Casentini, Caudill, Cavagli{\`a}, Cavalier, Cavalieri, Cella, Cepeda,
  Cerd{\'a}-Dur{\'a}n, Cerretani, Cesarini, Chamberlin, Chan, Chao, Charlton,
  Chase, Chassande-Mottin, Chatterjee, Chatziioannou, Cheeseboro, Chen, Chen,
  Chen, Cheng, Chia, Chincarini, Chiummo, Chmiel, Cho, Cho, Chow, Christensen,
  Chu, Chua, Chua, Chung, Chung, Ciani, Ciolfi, Cirelli, Cirone, Clara, Clark,
  Clearwater, Cleva, Cocchieri, Coccia, Cohadon, Cohen, Colla, Collette,
  Cominsky, Constancio, Conti, Cooper, Corban, Corbitt, Cordero-Carri{\'o}n,
  Corley, Cornish, Corsi, Cortese, Costa, Coughlin, Coughlin, Coulon,
  Countryman, Couvares, Covas, Cowan, Coward, Cowart, Coyne, Coyne, Creighton,
  Creighton, Cripe, Crowder, Cullen, Cumming, Cunningham, Cuoco, Dal~Canton,
  D{\'a}lya, Danilishin, D'Antonio, Danzmann, Dasgupta, Da~Silva~Costa,
  Dattilo, Dave, Davier, Davis, Daw, Day, De, DeBra, Degallaix, De~Laurentis,
  Del{\'e}glise, Del~Pozzo, Demos, Denker, Dent, De~Pietri, Dergachev, De~Rosa,
  DeRosa, De~Rossi, DeSalvo, de~Varona, Devenson, Dhurandhar, D{\'{\i}}az,
  Di~Fiore, Di~Giovanni, Di~Girolamo, Di~Lieto, Di~Pace, Di~Palma, Di~Renzo,
  Doctor, Dolique, Donovan, Dooley, Doravari, Dorrington, Douglas,
  Dovale~{\'A}lvarez, Downes, Drago, Dreissigacker, Driggers, Du, Ducrot,
  Dupej, Dwyer, Edo, Edwards, Effler, Eggenstein, Ehrens, Eichholz, Eikenberry,
  Eisenstein, Essick, Estevez, Etienne, Etzel, Evans, Evans, Factourovich,
  Fafone, Fair, Fairhurst, Fan, Farinon, Farr, Farr, Fauchon-Jones, Favata,
  Fays, Fee, Fehrmann, Feicht, Fejer, Fernandez-Galiana, Ferrante, Ferreira,
  Ferrini, Fidecaro, Finstad, Fiori, Fiorucci, Fishbach, Fisher, Fitz-Axen,
  Flaminio, Fletcher, Fong, Font, Forsyth, Forsyth, Fournier, Frasca, Frasconi,
  Frei, Freise, Frey, Frey, Fries, Fritschel, Frolov, Fulda, Fyffe, Gabbard,
  Gadre, Gaebel, Gair, Gammaitoni, Ganija, Gaonkar, Garcia-Quiros, Garufi,
  Gateley, Gaudio, Gaur, Gayathri, Gehrels, Gemme, Genin, Gennai, George,
  George, Gergely, Germain, Ghonge, Ghosh, Ghosh, Ghosh, Giaime, Giardina,
  Giazotto, Gill, Glover, Goetz, Goetz, Gomes, Goncharov, Gonz{\'a}lez,
  Gonzalez~Castro, Gopakumar, Gorodetsky, Gossan, Gosselin, Gouaty, Grado,
  Graef, Granata, Grant, Gras, Gray, Greco, Green, Gretarsson, Groot, Grote,
  Grunewald, Gruning, Guidi, Guo, Gupta, Gupta, Gushwa, Gustafson, Gustafson,
  Halim, Hall, Hall, Hamilton, Hammond, Haney, Hanke, Hanks, Hanna, Hannam,
  Hannuksela, Hanson, Hardwick, Harms, Harry, Harry, Hart, Haster, Haughian,
  Healy, Heidmann, Heintze, Heitmann, Hello, Hemming, Hendry, Heng, Hennig,
  Heptonstall, Heurs, Hild, Hinderer, Hoak, Hofman, Holt, Holz, Hopkins, Horst,
  Hough, Houston, Howell, Hreibi, Hu, Huerta, Huet, Hughey, Husa, Huttner,
  Huynh-Dinh, Indik, Inta, Intini, Isa, Isac, Isi, Iyer, Izumi, Jacqmin, Jani,
  Jaranowski, Jawahar, Jim{\'e}nez-Forteza, Johnson, Johnson-McDaniel, Jones,
  Jones, Jonker, Ju, Junker, Kalaghatgi, Kalogera, Kamai, Kandhasamy, Kang,
  Kanner, Kapadia, Karki, Karvinen, Kasprzack, Kastaun, Katolik, Katsavounidis,
  Katzman, Kaufer, Kawabe, K{\'e}f{\'e}lian, Keitel, Kemball, Kennedy, Kent,
  Key, Khalili, Khan, Khan, Khan, Khazanov, Kijbunchoo, Kim, Kim, Kim, Kim,
  Kim, Kim, Kimbrell, King, King, Kinley-Hanlon, Kirchhoff, Kissel, Kleybolte,
  Klimenko, Knowles, Koch, Koehlenbeck, Koley, Kondrashov, Kontos, Korobko,
  Korth, Kowalska, Kozak, Kr{\"a}mer, Kringel, Krishnan, Kr{\'o}lak, Kuehn,
  Kumar, Kumar, Kumar, Kuo, Kutynia, Kwang, Lackey, Lai, Landry, Lang, Lange,
  Lantz, Lanza, Lartaux-Vollard, Lasky, Laxen, Lazzarini, Lazzaro, Leaci,
  Leavey, Lee, Lee, Lee, Lee, Lee, Lehmann, Lenon, Leonardi, Leroy, Letendre,
  Levin, Li, Linker, Littenberg, Liu, Lo, Lockerbie, London, Lord, Lorenzini,
  Loriette, Lormand, Losurdo, Lough, Lousto, Lovelace, L{\"u}ck, Lumaca,
  Lundgren, Lynch, Ma, Macas, Macfoy, Machenschalk, MacInnis, Macleod,
  Maga{\~n}a~Hernandez, Maga{\~n}a-Sandoval, Maga{\~n}a~Zertuche, Magee,
  Majorana, Maksimovic, Man, Mandic, Mangano, Mansell, Manske, Mantovani,
  Marchesoni, Marion, M{\'a}rka, M{\'a}rka, Markakis, Markosyan, Markowitz,
  Maros, Marquina, Martelli, Martellini, Martin, Martin, Martynov, Mason,
  Massera, Masserot, Massinger, Masso-Reid, Mastrogiovanni, Matas, Matichard,
  Matone, Mavalvala, Mazumder, McCarthy, McClelland, McCormick, McCuller,
  McGuire, McIntyre, McIver, McManus, McNeill, McRae, McWilliams, Meacher,
  Meadors, Mehmet, Meidam, Mejuto-Villa, Melatos, Mendell, Mercer, Merilh,
  Merzougui, Meshkov, Messenger, Messick, Metzdorff, Meyers, Miao, Michel,
  Middleton, Mikhailov, Milano, Miller, Miller, Miller, Millhouse,
  Milovich-Goff, Minazzoli, Minenkov, Ming, Mishra, Mitra, Mitrofanov,
  Mitselmakher, Mittleman, Moffa, Moggi, Mogushi, Mohan, Mohapatra, Montani,
  Moore, Moraru, Moreno, Morriss, Mours, Mow-Lowry, Mueller, Muir, Mukherjee,
  Mukherjee, Mukherjee, Mukund, Mullavey, Munch, Mu{\~n}iz, Muratore, Murray,
  Napier, Nardecchia, Naticchioni, Nayak, Neilson, Nelemans, Nelson, Nery,
  Neunzert, Nevin, Newport, Newton, Ng, Nguyen, Nichols, Nielsen, Nissanke,
  Nitz, Noack, Nocera, Nolting, North, Nuttall, Oberling, O'Dea, Ogin, Oh, Oh,
  Ohme, Okada, Oliver, Oppermann, Oram, O'Reilly, Ormiston, Ortega,
  O'Shaughnessy, Ossokine, Ottaway, Overmier, Owen, Pace, Page, Page, Pai, Pai,
  Palamos, Palashov, Palomba, Pal-Singh, Pan, Pan, Pang, Pang, Pankow,
  Pannarale, Pant, Paoletti, Paoli, Papa, Parida, Parker, Pascucci,
  Pasqualetti, Passaquieti, Passuello, Patil, Patricelli, Pearlstone, Pedraza,
  Pedurand, Pekowsky, Pele, Penn, Perez, Perreca, Perri, Pfeiffer, Phelps,
  Piccinni, Pichot, Piergiovanni, Pierro, Pillant, Pinard, Pinto, Pirello,
  Pitkin, Poe, Poggiani, Popolizio, Porter, Post, Powell, Prasad, Pratt,
  Pratten, Predoi, Prestegard, Prijatelj, Principe, Privitera, Prodi,
  Prokhorov, Puncken, Punturo, Puppo, P{\"u}rrer, Qi, Quetschke, Quintero,
  Quitzow-James, Raab, Rabeling, Radkins, Raffai, Raja, Rajan, Rajbhandari,
  Rakhmanov, Ramirez, Ramos-Buades, Rapagnani, Raymond, Razzano, Read,
  Regimbau, Rei, Reid, Reitze, Ren, Reyes, Ricci, Ricker, Rieger, Riles, Rizzo,
  Robertson, Robie, Robinet, Rocchi, Rolland, Rollins, Roma, Romano, Romel,
  Romie, Rosi{\'n}ska, Ross, Rowan, R{\"u}diger, Ruggi, Rutins, Ryan, Sachdev,
  Sadecki, Sadeghian, Sakellariadou, Salconi, Saleem, Salemi, Samajdar, Sammut,
  Sampson, Sanchez, Sanchez, Sanchis-Gual, Sandberg, Sanders, Sassolas,
  Sathyaprakash, Saulson, Sauter, Savage, Sawadsky, Schale, Scheel, Scheuer,
  Schmidt, Schmidt, Schnabel, Schofield, Sch{\"o}nbeck, Schreiber, Schuette,
  Schulte, Schutz, Schwalbe, Scott, Scott, Seidel, Sellers, Sengupta, Sentenac,
  Sequino, Sergeev, Shaddock, Shaffer, Shah, Shahriar, Shaner, Shao, Shapiro,
  Shawhan, Sheperd, Shoemaker, Shoemaker, Siellez, Siemens, Sieniawska, Sigg,
  Silva, Singer, Singh, Singhal, Sintes, Slagmolen, Smith, Smith, Smith,
  Somala, Son, Sonnenberg, Sorazu, Sorrentino, Souradeep, Spencer, Srivastava,
  Staats, Staley, Steinke, Steinlechner, Steinlechner, Steinmeyer, Stevenson,
  Stone, Stops, Strain, Stratta, Strigin, Strunk, Sturani, Stuver,
  Summerscales, Sun, Sunil, Suresh, Sutton, Swinkels, Szczepa{\'n}czyk, Tacca,
  Tait, Talbot, Talukder, Tanner, T{\'a}pai, Taracchini, Tasson, Taylor,
  Taylor, Tewari, Theeg, Thies, Thomas, Thomas, Thomas, Thorne, Thorne, Thrane,
  Tiwari, Tiwari, Tokmakov, Toland, Tonelli, Tornasi, Torres-Forn{\'e}, Torrie,
  T{\"o}yr{\"a}, Travasso, Traylor, Trinastic, Tringali, Trozzo, Tsang, Tse,
  Tso, Tsukada, Tsuna, Tuyenbayev, Ueno, Ugolini, Unnikrishnan, Urban, Usman,
  Vahlbruch, Vajente, Valdes, van Bakel, van Beuzekom, van~den Brand, Van
  Den~Broeck, Vander-Hyde, van~der Schaaf, van Heijningen, van Veggel, Vardaro,
  Varma, Vass, Vas{\'u}th, Vecchio, Vedovato, Veitch, Veitch, Venkateswara,
  Venugopalan, Verkindt, Vetrano, Vicer{\'e}, Viets, Vinciguerra, Vine, Vinet,
  Vitale, Vo, Vocca, Vorvick, Vyatchanin, Wade, Wade, Wade, Walet, Walker,
  Wallace, Walsh, Wang, Wang, Wang, Wang, Wang, Ward, Warner, Was, Watchi,
  Weaver, Wei, Weinert, Weinstein, Weiss, Wen, Wessel, We{\ss}els, Westerweck,
  Westphal, Wette, Whelan, Whitcomb, Whiting, Whittle, Wilken, Williams,
  Williams, Williamson, Willis, Willke, Wimmer, Winkler, Wipf, Wittel, Woan,
  Woehler, Wofford, Wong, Worden, Wright, Wu, Wysocki, Xiao, Yamamoto, Yancey,
  Yang, Yap, Yazback, Yu, Yu, Yvert, Zadro{\.z}ny, Zanolin, Zelenova, Zendri,
  Zevin, Zhang, Zhang, Zhang, Zhang, Zhao, Zhou, Zhou, Zhu, Zhu, Zimmerman,
  Zucker, Zweizig, Collaboration, Collaboration, Burns, Veres, Kocevski,
  Racusin, Goldstein, Connaughton, Briggs, Blackburn, Hamburg, Hui, von
  Kienlin, McEnery, Preece, Wilson-Hodge, Bissaldi, Cleveland, Gibby, Giles,
  Kippen, McBreen, Meegan, Paciesas, Poolakkil, Roberts, Stanbro, Gamma-ray
  Burst~Monitor, Savchenko, Ferrigno, Kuulkers, Bazzano, Bozzo, Brandt,
  Chenevez, Courvoisier, Diehl, Domingo, Hanlon, Jourdain, Laurent, Lebrun,
  Lutovinov, Mereghetti, Natalucci, Rodi, Roques, Sunyaev, Ubertini, \&
  (INTEGRAL}]{Abbott2017}
Abbott, B.~P., Abbott, R., Abbott, T.~D., {et~al.} 2017, \apjl, 848, L13,
  \dodoi{10.3847/2041-8213/aa920c}

\bibitem[{Aloy {et~al.}(2005)Aloy, Janka, \& M{\"u}ller}]{Aloy_et_al_2005}
Aloy, M.~A., Janka, H.-T., \& M{\"u}ller, E. 2005, \aap, 436, 273,
  \dodoi{10.1051/0004-6361:20041865}

\bibitem[{Baiotti \& Rezzolla(2017)}]{BaiottiRezolla2016}
Baiotti, L., \& Rezzolla, L. 2017, Reports on Progress in Physics, 80, 096901,
  \dodoi{10.1088/1361-6633/aa67bb}

\bibitem[{Balbus \& Hawley(1991)}]{Balbus_Hawley_1991}
Balbus, S.~A., \& Hawley, J.~F. 1991, \apj, 376, 214, \dodoi{10.1086/170270}

\bibitem[{{Baring} \& {Harding}(1997)}]{Baring_1997}
{Baring}, M.~G., \& {Harding}, A.~K. 1997, \apj, 491, 663,
  \dodoi{10.1086/304982}

\bibitem[{Beckwith {et~al.}(2008)Beckwith, Hawley, \& Krolik}]{Beckwith_2008}
Beckwith, K., Hawley, J.~F., \& Krolik, J.~H. 2008, \apj, 678, 1180,
  \dodoi{10.1086/533492}

\bibitem[{{Bisnovatyi-Kogan} \& {Ruzmaikin}(1976)}]{Bisnovatyi_1976}
{Bisnovatyi-Kogan}, G.~S., \& {Ruzmaikin}, A.~A. 1976, \apss, 42, 401,
  \dodoi{10.1007/BF01225967}

\bibitem[{Blandford \& Znajek(1977)}]{Blandoford_Znajek_1977}
Blandford, R.~D., \& Znajek, R.~L. 1977, \mnras, 179, 433,
  \dodoi{10.1093/mnras/179.3.433}

\bibitem[{Chakrabarti(1985)}]{Chakrabarti1985}
Chakrabarti, S.~K. 1985, \apj, 288, 1, \dodoi{10.1086/162755}

\bibitem[{{Eichler} {et~al.}(1989){Eichler}, {Livio}, {Piran}, \&
  {Schramm}}]{Eichler_Livio_Piran_schramm_1989_Nature}
{Eichler}, D., {Livio}, M., {Piran}, T., \& {Schramm}, D.~N. 1989, \nat, 340,
  126, \dodoi{10.1038/340126a0}

\bibitem[{{Fern{\'a}ndez} {et~al.}(2018){Fern{\'a}ndez}, {Tchekhovskoy},
  {Quataert}, {Foucart}, \& {Kasen}}]{2018MNRAS.tmp.2798F}
{Fern{\'a}ndez}, R., {Tchekhovskoy}, A., {Quataert}, E., {Foucart}, F., \&
  {Kasen}, D. 2018, \mnras, \dodoi{10.1093/mnras/sty2932}

\bibitem[{{Ferrari} {et~al.}(2010){Ferrari}, {Gualtieri}, \&
  {Pannarale}}]{Ferrari_2010}
{Ferrari}, V., {Gualtieri}, L., \& {Pannarale}, F. 2010, \prd, 81, 064026,
  \dodoi{10.1103/PhysRevD.81.064026}

\bibitem[{{Fishbone} \& {Moncrief}(1976)}]{Fishbone_Moncrief_1976_ApJ}
{Fishbone}, L.~G., \& {Moncrief}, V. 1976, \apj, 207, 962,
  \dodoi{10.1086/154565}

\bibitem[{Fong {et~al.}(2015)Fong, Berger, Margutti, \&
  Zauderer}]{Fongetal2015}
Fong, W., Berger, E., Margutti, R., \& Zauderer, B.~A. 2015, \apj, 815, 102,
  \dodoi{10.1088/0004-637X/815/2/102}

\bibitem[{Font {et~al.}(1999)Font, Ib{\'a}{\~n}ez, \&
  Papadopoulos}]{Fontetal1999}
Font, J.~A., Ib{\'a}{\~n}ez, J.~M., \& Papadopoulos, P. 1999, \mnras, 305, 920,
  \dodoi{10.1046/j.1365-8711.1999.02459.x}

\bibitem[{Gammie(2004)}]{Gammie_2004}
Gammie, C.~F. 2004, \apj, 614, 309, \dodoi{10.1086/423443}

\bibitem[{{Gammie} {et~al.}(2003){Gammie}, {McKinney}, \&
  {T{\'o}th}}]{Gammie_2003}
{Gammie}, C.~F., {McKinney}, J.~C., \& {T{\'o}th}, G. 2003, \apj, 589, 444,
  \dodoi{10.1086/374594}

\bibitem[{Garofalo \& Meier(2010)}]{Garofaloetal2010}
Garofalo, D., \& Meier, D.~L. 2010, \mnras, 406, 2047,
  \dodoi{10.1111/j.1365-2966.2010.16815.x}

\bibitem[{Granot {et~al.}(2017)Granot, Guetta, \& Gill}]{Granot2017}
Granot, J., Guetta, D., \& Gill, R. 2017, \apjl, 850, L24,
  \dodoi{10.3847/2041-8213/aa991d}

\bibitem[{Granot {et~al.}(2015)Granot, Piran, Bromberg, Racusin, \&
  Daigne}]{Granot2015}
Granot, J., Piran, T., Bromberg, O., Racusin, J.~L., \& Daigne, F. 2015, \ssr,
  191, 471, \dodoi{10.1007/s11214-015-0191-6}

\bibitem[{{Guan} \& {Gammie}(2008)}]{2008ApJS..174..145G}
{Guan}, X., \& {Gammie}, C.~F. 2008, \apjs, 174, 145, \dodoi{10.1086/521147}

\bibitem[{Hawley {et~al.}(2011)Hawley, Guan, \& Krolik}]{Hawley_2011}
Hawley, J.~F., Guan, X., \& Krolik, J.~H. 2011, \apj, 738, 84,
  \dodoi{10.1088/0004-637X/738/1/84}

\bibitem[{Jackson(1998)}]{jackson_classical_electrodynamic}
Jackson, J.~D. 1998, Classical Electrodynamics, 3rd Edition (Wiley), 832

\bibitem[{Janiuk(2017)}]{Janiuk2017}
Janiuk, A. 2017, \apj, 837, 39, \dodoi{10.3847/1538-4357/aa5f16}

\bibitem[{Janiuk {et~al.}(2013)Janiuk, Mioduszewski, \&
  Moscibrodzka}]{Janiuk2013}
Janiuk, A., Mioduszewski, P., \& Moscibrodzka, M. 2013, \apj, 776, 105,
  \dodoi{10.1088/0004-637X/776/2/105}

\bibitem[{Just {et~al.}(2016)Just, Obergaulinger, Janka, Bauswein, \&
  Schwarz}]{Justetal2016}
Just, O., Obergaulinger, M., Janka, H.-T., Bauswein, A., \& Schwarz, N. 2016,
  \apjl, 816, L30, \dodoi{10.3847/2041-8205/816/2/L30}

\bibitem[{{Kathirgamaraju} {et~al.}(2018{\natexlab{a}}){Kathirgamaraju},
  {Barniol Duran}, \& {Giannios}}]{2018MNRAS.473L.121K}
{Kathirgamaraju}, A., {Barniol Duran}, R., \& {Giannios}, D.
  2018{\natexlab{a}}, \mnras, 473, L121, \dodoi{10.1093/mnrasl/slx175}

\bibitem[{{Kathirgamaraju} {et~al.}(2018{\natexlab{b}}){Kathirgamaraju},
  {Tchekhovskoy}, {Giannios}, \& {Barniol Duran}}]{2018arXiv180905099K}
{Kathirgamaraju}, A., {Tchekhovskoy}, A., {Giannios}, D., \& {Barniol Duran},
  R. 2018{\natexlab{b}}, ArXiv e-prints.
\newblock \doarXiv{1809.05099}

\bibitem[{Komissarov(2001)}]{Komissarov2001}
Komissarov, S.~S. 2001, \mnras, 326, L41,
  \dodoi{10.1046/j.1365-8711.2001.04863.x}

\bibitem[{Komissarov(2009)}]{Komissarov2009}
---. 2009, Journ of Korean Phys. Soc., 54, 2503, \dodoi{10.3938/jkps.54.2503}

\bibitem[{{Kumar} {et~al.}(2017){Kumar}, {P{\"u}rrer}, \&
  {Pfeiffer}}]{Kumar_2017}
{Kumar}, P., {P{\"u}rrer}, M., \& {Pfeiffer}, H.~P. 2017, \prd, 95, 044039,
  \dodoi{10.1103/PhysRevD.95.044039}

\bibitem[{Lazzati {et~al.}(2017)Lazzati, Deich, Morsony, \&
  Workman}]{Lazzati2017}
Lazzati, D., Deich, A., Morsony, B.~J., \& Workman, J.~C. 2017, \mnras, 471,
  1652, \dodoi{10.1093/mnras/stx1683}

\bibitem[{{Lei} {et~al.}(2013){Lei}, {Zhang}, \& {Liang}}]{Lei_2013}
{Lei}, W.-H., {Zhang}, B., \& {Liang}, E.-W. 2013, \apj, 765, 125,
  \dodoi{10.1088/0004-637X/765/2/125}

\bibitem[{Liska {et~al.}(2018)Liska, Hesp, Tchekhovskoy, Ingram, van~der Klis,
  \& Markoff}]{Liska2018}
Liska, M., Hesp, C., Tchekhovskoy, A., {et~al.} 2018, \mnras, 474, L81,
  \dodoi{10.1093/mnrasl/slx174}

\bibitem[{Liu {et~al.}(2015)Liu, Hou, Xue, \& Gu}]{Liu2015}
Liu, T., Hou, S.-J., Xue, L., \& Gu, W.-M. 2015, \apjs, 218, 12,
  \dodoi{10.1088/0067-0049/218/1/12}

\bibitem[{{Lyutikov} \& {Blandford}(2003)}]{2003astro.ph.12347L}
{Lyutikov}, M., \& {Blandford}, R. 2003, ArXiv Astrophysics e-prints

\bibitem[{MacLachlan {et~al.}(2013)MacLachlan, Shenoy, Sonbas, Dhuga, Cobb,
  Ukwatta, Morris, Eskandarian, Maximon, \& Parke}]{MacLachlan2013}
MacLachlan, G.~A., Shenoy, A., Sonbas, E., {et~al.} 2013, \mnras, 432, 857,
  \dodoi{10.1093/mnras/stt241}

\bibitem[{Margalit \& Metzger(2017)}]{MargalitMetgzer2017}
Margalit, B., \& Metzger, B.~D. 2017, \apjl, 850, L19,
  \dodoi{10.3847/2041-8213/aa991c}

\bibitem[{Margalit {et~al.}(2015)Margalit, Metzger, \&
  Beloborodov}]{Margalit2015}
Margalit, B., Metzger, B.~D., \& Beloborodov, A.~M. 2015, Phys. Rev. Lett.,
  115, 171101, \dodoi{10.1103/PhysRevLett.115.171101}

\bibitem[{{McKinney} {et~al.}(2012){McKinney}, {Tchekhovskoy}, \&
  {Blandford}}]{McKinney_2012}
{McKinney}, J.~C., {Tchekhovskoy}, A., \& {Blandford}, R.~D. 2012, \mnras, 423,
  3083, \dodoi{10.1111/j.1365-2966.2012.21074.x}

\bibitem[{{Metzger} \& {Fern{\'a}ndez}(2014)}]{2014MNRAS.441.3444M}
{Metzger}, B.~D., \& {Fern{\'a}ndez}, R. 2014, \mnras, 441, 3444,
  \dodoi{10.1093/mnras/stu802}

\bibitem[{Mochkovitch {et~al.}(1993)Mochkovitch, Hernanz, Isern, \&
  Martin}]{Mochkovitch_et_al_1993}
Mochkovitch, R., Hernanz, M., Isern, J., \& Martin, X. 1993, \nat, 361, 236,
  \dodoi{10.1038/361236a0}

\bibitem[{Murguia-Berthier {et~al.}(2017)Murguia-Berthier, Ramirez-Ruiz,
  Kilpatrick, Foley, Kasen, Lee, Piro, Coulter, Drout, Madore, Shappee, Pan,
  Prochaska, Rest, Rojas-Bravo, Siebert, \& Simon}]{Murguia_Berthier2017}
Murguia-Berthier, A., Ramirez-Ruiz, E., Kilpatrick, C.~D., {et~al.} 2017,
  \apjl, 848, L34, \dodoi{10.3847/2041-8213/aa91b3}

\bibitem[{{Narayan} {et~al.}(1992){Narayan}, {Paczynski}, \&
  {Piran}}]{Narayanetal1992}
{Narayan}, R., {Paczynski}, B., \& {Piran}, T. 1992, \apjl, 395, L83,
  \dodoi{10.1086/186493}

\bibitem[{Narayan {et~al.}(2012)Narayan, S{\"A dowski}, Penna, \&
  Kulkarni}]{Narayan_et_al_2012}
Narayan, R., S{\"A dowski}, A., Penna, R.~F., \& Kulkarni, A.~K. 2012, \mnras,
  426, 3241, \dodoi{10.1111/j.1365-2966.2012.22002.x}

\bibitem[{{Noble} {et~al.}(2006){Noble}, {Gammie}, {McKinney}, \& {Del
  Zanna}}]{Noble_et_all_2006}
{Noble}, S.~C., {Gammie}, C.~F., {McKinney}, J.~C., \& {Del Zanna}, L. 2006,
  \apj, 641, 626, \dodoi{10.1086/500349}

\bibitem[{Noble {et~al.}(2010)Noble, Krolik, \& Hawley}]{Noble_2010}
Noble, S.~C., Krolik, J.~H., \& Hawley, J.~F. 2010, \apj, 711, 959,
  \dodoi{10.1088/0004-637X/711/2/959}

\bibitem[{Novikov \& Thorne(1973)}]{Novikov_Thorne_1973}
Novikov, I.~D., \& Thorne, K.~S. 1973, in Black Holes (Les Astres Occlus), ed.
  C.~{Dewitt} \& B.~S. {Dewitt}, 343--450

\bibitem[{{Paczynski}(1986)}]{Paczynski_1986}
{Paczynski}, B. 1986, \apjl, 308, L43, \dodoi{10.1086/184740}

\bibitem[{{Paczynski}(1991)}]{Paczynski_1991_Acta_Astronomica}
---. 1991, \actaa, 41, 257

\bibitem[{{Paschalidis}(2017)}]{Paschalidis_2017}
{Paschalidis}, V. 2017, Classical and Quantum Gravity, 34, 084002,
  \dodoi{10.1088/1361-6382/aa61ce}

\bibitem[{{Paschalidis} {et~al.}(2013){Paschalidis}, {Etienne}, \&
  {Shapiro}}]{Paschalidis2013}
{Paschalidis}, V., {Etienne}, Z.~B., \& {Shapiro}, S.~L. 2013, \prd, 88,
  021504, \dodoi{10.1103/PhysRevD.88.021504}

\bibitem[{Paschalidis {et~al.}(2015)Paschalidis, Ruiz, \&
  Shapiro}]{Paschalidis_Ruiz_Shapiro_2015}
Paschalidis, V., Ruiz, M., \& Shapiro, S.~L. 2015, \apjl, 806, L14,
  \dodoi{10.1088/2041-8205/806/1/L14}

\bibitem[{Penna {et~al.}(2013)Penna, Kulkarni, \& Narayan}]{Pennaetal2013}
Penna, R.~F., Kulkarni, A., \& Narayan, R. 2013, \aap, 559, A116,
  \dodoi{10.1051/0004-6361/201219666}

\bibitem[{{Penna} {et~al.}(2013){Penna}, {Narayan}, \& {S{\c a}dowski}}]{Penna}
{Penna}, R.~F., {Narayan}, R., \& {S{\c a}dowski}, A. 2013, \mnras, 436, 3741,
  \dodoi{10.1093/mnras/stt1860}

\bibitem[{{Perego} {et~al.}(2014){Perego}, {Rosswog}, {Cabez{\'o}n},
  {Korobkin}, {K{\"a}ppeli}, {Arcones}, \&
  {Liebend{\"o}rfer}}]{2014MNRAS.443.3134P}
{Perego}, A., {Rosswog}, S., {Cabez{\'o}n}, R.~M., {et~al.} 2014, \mnras, 443,
  3134, \dodoi{10.1093/mnras/stu1352}

\bibitem[{{Perna} {et~al.}(2018){Perna}, {Lazzati}, \&
  {Cantiello}}]{pernaetal2018}
{Perna}, R., {Lazzati}, D., \& {Cantiello}, M. 2018, \apj, 859, 48,
  \dodoi{10.3847/1538-4357/aabcc1}

\bibitem[{Rees \& Meszaros(1994)}]{ReesMeszaros1994}
Rees, M.~J., \& Meszaros, P. 1994, \apjl, 430, L93, \dodoi{10.1086/187446}

\bibitem[{Rezzolla {et~al.}(2011)Rezzolla, Giacomazzo, Baiotti, Granot,
  Kouveliotou, \& Aloy}]{Rezzolla_et_al_2011}
Rezzolla, L., Giacomazzo, B., Baiotti, L., {et~al.} 2011, \apjl, 732, L6,
  \dodoi{10.1088/2041-8205/732/1/L6}

\bibitem[{Rhoads(1999)}]{Rhoads1999}
Rhoads, J.~E. 1999, \apj, 525, 737, \dodoi{10.1086/307907}

\bibitem[{Ruiz {et~al.}(2016)Ruiz, Lang, Paschalidis, \& Shapiro}]{Ruiz2016}
Ruiz, M., Lang, R.~N., Paschalidis, V., \& Shapiro, S.~L. 2016, \apjl, 824, L6,
  \dodoi{10.3847/2041-8205/824/1/L6}

\bibitem[{Sari {et~al.}(1999)Sari, Piran, \& Halpern}]{Sarietal1999}
Sari, R., Piran, T., \& Halpern, J.~P. 1999, \apjl, 519, L17,
  \dodoi{10.1086/312109}

\bibitem[{{Shemi} \& {Piran}(1990)}]{Shemi_1990}
{Shemi}, A., \& {Piran}, T. 1990, \apjl, 365, L55, \dodoi{10.1086/185887}

\bibitem[{Shibata {et~al.}(2000)Shibata, Baumgarte, \& Shapiro}]{Shibata2000}
Shibata, M., Baumgarte, T.~W., \& Shapiro, S.~L. 2000, \prd, 61, 044012,
  \dodoi{10.1103/PhysRevD.61.044012}

\bibitem[{Tchekhovskoy \& Giannios(2015)}]{Tchekhovskoy_Giannios2015}
Tchekhovskoy, A., \& Giannios, D. 2015, \mnras, 447, 327,
  \dodoi{10.1093/mnras/stu2229}

\bibitem[{Vlahakis \& K{\"o}nigl(2003)}]{Vlahakis_2003a}
Vlahakis, N., \& K{\"o}nigl, A. 2003, \apj, 596, 1080, \dodoi{10.1086/378226}

\bibitem[{{Woosley}(1993)}]{woosley1993}
{Woosley}, S.~E. 1993, \apj, 405, 273, \dodoi{10.1086/172359}

\bibitem[{{Woosley} \& {Heger}(2006)}]{woosleyheger2006}
{Woosley}, S.~E., \& {Heger}, A. 2006, \apj, 637, 914, \dodoi{10.1086/498500}

\bibitem[{Yang {et~al.}(2015)Yang, Zhang, \& Lehner}]{Yang_2015}
Yang, H., Zhang, F., \& Lehner, L. 2015, \prd, 91, 124055,
  \dodoi{10.1103/PhysRevD.91.124055}

\bibitem[{Zhang {et~al.}(2017)Zhang, Zhang, Sun, Lei, Gao, Li, Shao, Zhao, Hu,
  L{\"u}, Wu, Fan, Wang, Castro-Tirado, Zhang, Yu, Cao, \&
  Liang}]{Zhang_et_al_2017arXiV}
Zhang, B.-B., Zhang, B., Sun, H., {et~al.} 2017, ArXiv e-prints.
\newblock \doarXiv{1710.05851}

\end{thebibliography}

\end{document}